\documentclass{aa}  

\usepackage{graphicx}
\usepackage{array}
\usepackage{tabularx}
\usepackage{amsmath}
\usepackage{multirow}
\usepackage{txfonts}
\usepackage{natbib}
\bibpunct{(}{)}{;}{a}{}{,} 
\usepackage{units}
\usepackage{comment}
\usepackage[normalem]{ulem}
\usepackage{placeins}
\newcolumntype{L}[1]{>{\raggedright\arraybackslash}p{#1}} 
\newcolumntype{C}[1]{>{\centering\arraybackslash}p{#1}} 
\newcolumntype{R}[1]{>{\raggedleft\arraybackslash}p{#1}} 
\newcommand{\msun}{M$_{\sun}$ }
\newcommand{\inv}[1]{#1$^{-1}$}

\usepackage{color}

\begin{document}

   \title{Core-collapse supernovae in the hall of mirrors.\\ A three-dimensional code-comparison project}

   \author{Rub\'en M. Cabez\'on\inst{1}
          \and Kuo-Chuan Pan\inst{2,3}
          \and Matthias Liebend\"orfer\inst{4}
          \and Takami Kuroda\inst{5}
          \and Kevin Ebinger\inst{6}
          \and Oliver Heinimann\inst{4}
          \and Albino Perego\inst{7}
          \and Friedrich-Karl Thielemann\inst{4,6}
          }
   \institute{Scientific Computing Core (sciCORE), Universit\"at Basel, Klingelbergstrasse 61, 4056 Basel, Switzerland\\
                        \email{ruben.cabezon@unibas.ch}
         \and Department of Physics and Institute of Astronomy, National Tsing Hua University, 30013 Hsinchu, Taiwan\\
                        \email{kpan@phys.nthu.edu.tw}
         \and Department of Physics and Astronomy, Michigan State University, 567 Wilson Rd., East Lansing, MI 48824, USA                        
         \and Departement Physik, Uni versit\"at Basel, Klingelbergstrasse 82, 4056 Basel, Switzerland\\
                        \email{matthias.liebendoerfer@unibas.ch}
         \and Institute f\"ur Kernphysik, TU Darmstadt, Schlossgartenstra\ss e 2, 64289 Darmstadt, Germany\\
                        \email{takami.kuroda@physik.tu-darmstadt.de}
         \and GSI Helmholtzzentrum f\"ur Schwerionenforschung, Planckstra\ss e 1, 64291 Darmstadt, Germany
         \and Istituto Nazionale di Fisica Nucleare, Sezione di Milano Biocca, Piazza della Scienza 3, 20126 Milano, Italy
             }

   \date{Received June 21, 2018}
 
  \abstract
   {Modeling core-collapse supernovae (SNe) with neutrino transport in three dimensions (3D) requires tremendous computing resources and some level of approximation. We present a first comparison study of core-collapse SNe in 3D with different physics approximations and hydrodynamics codes.}
   {The objective of this work is to assess the impact of the hydrodynamics code, approximations for the neutrino, gravity treatments, and rotation on the simulation of core-collapse SNe in 3D.}
   {We use four different hydrodynamics codes in this work (ELEPHANT, FLASH, fGR1, and SPHYNX) in combination with two different neutrino treatments, the isotropic diffusion source approximation (IDSA) and two-moment M1, and three different gravity treatments (Newtonian, 1D General Relativity correction, and full General Relativity). Additional parameters discussed in this study are the inclusion of neutrino-electron scattering via a parametrized deleptonization and the influence of rotation.}
   {The four codes compared in this work include Eulerian and fully Lagrangian (smoothed particle hydrodynamics) codes for the first time. They show agreement in the overall evolution of the collapse phase and early post-bounce within the range of 10\% (20\% in some cases). The comparison of the different neutrino treatments highlights the need to further investigate the antineutrino luminosities in IDSA, which tend to be relatively high. We also demonstrate the requirement for a more detailed heavy-lepton neutrino leakage. When comparing with a full General Relativity code, including an M1 transport method, we confirm the influence of neutrino-electron scattering during the collapse phase, which is adequately captured by the parametrized deleptonization scheme. Also, the effective general relativistic potential reproduces the overall dynamic evolution correctly in all Newtonian codes. Additionally, we verify that rotation aids the shock expansion and estimate the overall angular momentum losses for each code in rotating scenarios.}
   {}


   \titlerunning{CCSNe in the hall of mirrors. A 3D code-comparison project}
   \maketitle
%

\section{Introduction}
\label{introduction}
Core-collapse supernovae (CCSNe) are violent stellar explosions of massive stars with masses $\gtrsim 8$~\msun at the end of their stellar evolution. 
Recent observations provide many details on CCSNe and their remnants, for example~light curves, spectra, 
large-scale asymmetry, and mixing \citep{aschenbach1995, delaney2010, milisavljevic2015, boggs2015}. 
However, there are many open questions in CCSN theory and in particular the explosion mechanism still produces explosions in numerical simulations that are sub-energetic when compared to observations (see recent reviews of \citealt{foglizzo2015, mezzacappa2015, janka2016, mirizzi2016, muller2016b, richers2017, burrows2018}).

While it is generally believed that the neutrino-driven convection and multi-dimensional hydrodynamics 
instabilities are key ingredients for successful explosions, 
modeling CCSNe in three dimensions (3D) with Boltzmann transport is still numerically expensive 
\citep{sumiyoshi2012, sumiyoshi2015, nagakura2018}
and the spatial resolution is not enough to resolve hydrodynamic instabilities and 
turbulence with current supercomputers \citep{couch2015, radice2016, radice2018}. 
Therefore, approximate schemes for the transport of neutrinos in multiple dimensions are still necessary, 
although they have to be sophisticated enough to capture the essential macro- and micro-physics.   

Several two-dimensional (2D) simulations with different approximate schemes have recently been performed and investigated, however the results among different groups are less consistent than expected. 
For instance, the same set of simulations of the 12, 15, 20, and 25~\msun progenitors shows clear differences  in the shock radius evolutions during the first hundred milliseconds post-bounce between the CHIMERA code from the Oak Ridge group \citep{bruenn2013, bruenn2016} and the PROMETHEUS-VERTEX code from the Garching group \citep{summa2016}, where the former uses flux-limited diffusion for the neutrino transport and the latter uses two-moment transport with a variable Eddington factor closure.
The Oak Ridge group and the Garching group use the so-called ray-by-ray-plus approach, 
which allocates hundreds of radial rays in the
computational domain. On each ray, an independent spherically symmetric
neutrino transport problem is solved, while in the optically thick regime, the neutrinos are additionally
allowed to laterally advect with the fluid between the rays. The CASTRO and FORNAX simulations by the Princeton group, with flux-limited diffusion \citep{dolence2015} 
and two-moment transport \citep{skinner2016, vartanyan2018} do not adopt the ray-by-ray approach 
and they find that the "explodability" is sensitive to the neutrino transport scheme and the detailed opacities \citep{burrows2018}. 

\cite{liebendoerfer2001,muller2012,oconnor2018} 
report that the higher neutrino luminosities and root mean square ($rms$) energies in full general relativity (GR) or effective GR calculations 
favor stronger supernova (SN) explosions, 
and \cite{oconnor2018} explicitly show that the effective GR potential correction \citep{marek2006} could turn failed SNe into explosions in both 1D and 2D. 
In addition, the FLASH-M1 simulations by \cite{oconnor2018} give qualitatively similar results to those of the Garching group, but ignore inelastic neutrino scattering processes. 
Furthermore, the Newtonian calculations from the Japanese group \citep{suwa2016, nakamura2015} 
and the Basel group \citep{pan2016} with the isotropic diffusion source approximation 
(IDSA, \citealt{liebendoerfer2009}) also give different results on the shock radius evolution and explodability. 

It should be noted that not only are the progenitor stars, the microphysics - such as neutrino interactions during collapse and post-bounce -, and the nuclear equations of state (EOS)  key factors for the CCSN problem, but also
the neutrino transport schemes, the implementation of hydrodynamics, the dimensionality of the problem, and
the spatial and temporal resolution are crucial for reliable SN models.
Therefore, the complexity of the problem makes it difficult to conduct a detailed comparison of SN codes
among different groups, 
which requires  computationally demanding simulations to be re-run and large amounts of 3D data to be exchanged and analyzed. 

Core-collapse supernova comparisons of neutrino transport schemes were first performed for the collapse phase \citep{mezzacappa1993a, mezzacappa1993c} with a focus on neutrino-electron scattering (NES), and were later
continued to the post-bounce phase by the comparison of the  VERTEX and AGILE-BOLTZTRAN codes \citep{liebendoerfer2005b}.
This work has recently been extended to six CCSN codes by \cite{oconnor2018b}.  
The latter work found very consistent results, but the comparison was limited to spherically symmetric calculations. 
Recently, \cite{just2018} presented a detailed 1D and 2D code comparison between the ALCAR and VERTEX codes. 
They discuss the influence of the ray-by-ray implementation and the impact of several different physical inputs, 
such as neglecting the neutrino-electron scattering, velocity, and redshift effects. 
However, ALCAR and VERTEX codes share similar discretization schemes and the comparison is limited to 1D and 2D.
Due to the larger amount of computing time required for 3D simulations, 
these have only been done with a few progenitor models 
\citep{lentz2015, kuroda2016, melson2015a, melson2015b, takiwaki2012,takiwaki2016, roberts2016, kuroda2017, ott2017, chan2018, oconnor2018c, summa2018}. A comparison of 3D models has not yet been documented.
Nevertheless, the nature of CCSN is inherently 3D \citep{delaney2010, couch2013b, boggs2015, muller2015, muller2016, muller2017}. 

In this study, we present a detailed SN code comparison in 3D.
In particular, we compare the neutrino treatment with the IDSA 
in three different hydrodynamics codes: the uniform Cartesian grid code ELEPHANT \citep{kaeppeli2011,liebendoerfer2009}, 
the adaptive mesh refinement (AMR) code FLASH \citep{fryxell2000}, 
and the smoothed particle hydrodynamics (SPH) code SPHYNX \citep{cabezon2017}. 
All three codes share the same IDSA kernel and use the same set of neutrino interactions and opacities.
However, due to the different discretizations of space and time in the three hydrodynamics codes, the
coupling of the IDSA to the hydrodynamic quantities had to be developed individually for each code (see Sect. 2).
In addition, we compare the results of these three codes with a 3D full GR code with two-moment M1 transport:
fGR1 \citep{kuroda2016}. The 1D AGILE-BOLTZTRAN code with Boltzmann transport \citep{liebendoerfer2004} is also
used for reference calculations in spherical symmetry.
Nevertheless, the differences in the discretization of space and time
of the four codes in combination with the 3D degrees of freedom of the simulation data do not allow us
to attribute clearly distinguishable and quantifiable effects in the results to particular code features.
Rather we aim to quantify a band of uncertainty into which the results of 3D SN simulations may
fall if they use different approaches for the implementation of hydrodynamics and neutrino transport. Our four methods differ most with respect to the computational domain (mesh refinement, box in a sphere, Lagrangian particles), the choice and implementation of gravity (effective GR potential for the monopole term, full GR),
neutrino transport (IDSA approximation, M1 transport) and parallelization (MPI, OpenMP, or using graphics processing unit -GPU- acceleration).
The codes with full neutrino transport can switch on NES, while the codes based on the IDSA cannot.
The goal of this comparison is not to single out a "best code" to be used in the future (it might not
be one of our four), but to further promote an approach where different codes are simultaneously used for
the same physical problem. If the virtues and weaknesses of each code are well-known from previous
comparisons, their results can be weighted accordingly in an uncertainty-aware interpretation of the simulated
physics.

This paper is organized as follows: We describe the four SN codes and 
their implementations of neutrino radiation transport in the following section. 
In Section~3, we define the runs that are compared and summarize the simulation data. 
A detailed comparison of the most relevant physical magnitudes is presented in Section~4. In Section~5, we summarize our results and conclusions.

\section{Methods and implementation}
\label{sec:methods}

In the following section we present the four hydrodynamic codes investigated in this work. We provide an overview of their main characteristics focusing on their implementation of the Euler equations, gravity evaluation, and their coupling with the neutrino treatment.

\subsection{ELEPHANT}
\label{elephant}
The ELEPHANT code (ELegant and Efficient Parallel Hydrodynamics with
Approximate Neutrino Transport) models the collapse and post-bounce
evolution of the innermost region of a massive star in 3D. ELEPHANT 
is based on the 3D magneto-hydrodynamics code FISH \citep{kaeppeli2011}
and implements IDSA
for the neutrino transport \citep{liebendoerfer2009,berninger2013}. The
IDSA is based on two key ideas: firstly, it treats trapped neutrinos
and streaming neutrinos as if they were separate particle species
that are linked by a production/annihilation rate $\Sigma$. Secondly,
as with flux-limited diffusion, the transition between diffusion and
free streaming is handled by interpolation. The main strength of ELEPHANT
is its computational simplicity and efficiency. The downside of this code is an equidistant mesh that limits the computational domain to the
innermost $\sim\unit[400]{km}$ of the star. The evolution of the
outer layers of the progenitor star is handled by the code AGILE-IDSA
\citep{liebendoerfer2002,liebendoerfer2009}\footnote{\url{https://astro.physik.unibas.ch/agile-idsa}},
which provides spherically symmetric data for the accreting matter
at the boundary of the 3D computational domain.

With respect to the input physics, ELEPHANT only implements the basics:
the Lattimer-Swesty equation of state \citep{lattimer1991} with an
incompressibility parameter $K=\unit[220]{MeV}$ and the neutrino
interaction rates defined in \cite{bruenn1985}. In comparison to
full Boltzmann neutrino transport in spherical symmetry \citep{liebendoerfer2004},
we found that neutrino-electron scattering of electron flavor neutrinos
plays a significant role in the collapse phase, but becomes less significant during the post-bounce phase. 
This is consistent with the findings of \cite{bruenn1989}, \cite{myra1989}, and \cite{pan2016},
however the findings of \cite{lentz2012b} show that more modern electron capture rates on nuclei
outperform neutrino-electron scattering in the collapse phase.
During the collapse phase, we take the accelerated deleptonization
effectively into account by using a parameterized deleptonization scheme
\citep{liebendoerfer2005}. In the post-bounce phase, we use the IDSA
and neglect neutrino-electron scattering. The energy loss caused by
the emission of $\mu$/$\tau$-neutrinos is handled by a gray leakage scheme (Sect.~\ref{additionalphysics}).

The hydrodynamics part of ELEPHANT solves the following set of 3D
conservation equations with gravitational source terms:
\begin{eqnarray}
\frac{\partial\rho}{\partial t}+\frac{\partial}{\partial x_{j}}\left(\rho v_{j}\right) & = & 0\label{eq:hydro1},\\
\frac{\partial}{\partial t}\left(\rho v_{i}\right)+\frac{\partial}{\partial x_{j}}\left(\rho v_{i}v_{j}+b_{i}b_{j}+p\delta_{ij}\right) & = & -\rho\frac{\partial\varphi}{\partial x_{i}},\nonumber \\
\frac{\partial E}{\partial t}+\frac{\partial}{\partial x_{j}}\left[\left(E+p\right)v_{j}\right] & = & -\rho v_{i}\frac{\partial\varphi}{\partial x_{i}},\nonumber \\
\frac{\partial}{\partial t}\left(\rho Y_{e}\right)+\frac{\partial}{\partial x_{j}}\left(\rho Y_{e}v_{j}\right) & = & 0,\nonumber \\
\frac{\partial}{\partial t}\left(\rho Y_{\nu}^{t}\right)+\frac{\partial}{\partial x_{j}}\left(\rho Y_{\nu}^{t}v_{j}\right) & = & 0,\nonumber \\
\frac{\partial}{\partial t}\left[\left(\rho Z_{\nu}^{t}\right)^{\frac{3}{4}}\right]+\frac{\partial}{\partial x_{j}}\left[\left(\rho Z_{\nu}^{t}\right)^{\frac{3}{4}}v_{j}\right] & = & 0,\label{eq:entropy.advection}\\
\frac{\partial}{\partial x_{i}}\left(\frac{\partial\varphi}{\partial x_{i}}\right) & = & 4\pi G\rho.\label{eq:hydro7}
\end{eqnarray}
The calculated unknowns, as functions of time $t$ and space $x_{i}$,
are the baryonic mass density $\rho$, the velocity $v_{i}$, the
electron fraction $Y_{e}$, the gravitational potential $\varphi$,
and the temperature $T$, this latter entering the equations via the specific
internal energy $e\left(\rho,T,Y_{e}\right)$ and the total energy
$E=\rho e+\rho v^{2}/2$. The pressure $p\left(\rho,T,Y_{e}\right)$
is provided by the equation of state. We modify the gravitational
potential by general relativistic corrections following \cite{marek2006}
in order to obtain a more realistic compactness of the proto-neutron
star (PNS). The factor $G$ is the gravitational constant. As a part of the IDSA, the
hydrodynamics equations additionally include equations that advect
the trapped electron neutrino fractions $Y_{\nu}^{t}$ and a multiple
of the neutrino entropies, $\left(\rho Z_{\nu}^{t}\right)^{\frac{3}{4}}$,
where $Z_{\nu}$ represents the mean neutrino specific energy. Analogous
equations are solved for the corresponding quantities of the electron
antineutrinos. ELEPHANT additionally includes the operator split evolution
of a magnetic field $B_{i}=4\pi b_{i}$ in the hydrodynamics part,
\begin{eqnarray*}
\frac{\partial\vec{b}}{\partial t}-\vec{\nabla}\times\left(\vec{v}\times\vec{b}\right) & = & 0.
\end{eqnarray*}
However, magnetic fields were ignored in the present comparison because there
were no data to compare with.
The neutrino transport equation is approximated by the IDSA; it separates the neutrino number density into a trapped neutrino distribution
function $f_{\nu}^{t}$ and a streaming neutrino distribution function
$f_{\nu}^{s}$, and evolves these two components separately. Based
on $Y_{\nu}^{t}$ and $Z_{\nu}^{t}$ we first construct the neutrino
distribution functions $f_{\nu}^{t}$ assuming an equilibrium spectrum
for the trapped component. Then, the diffusion equation
\begin{eqnarray}
\frac{\partial f_{\nu}^{{\rm t}}}{c\partial t} & = & j_{\nu}-\left(j_{\nu}+\chi_{\nu}\right)f_{\nu}^{{\rm t}}-\Sigma_{\nu},\label{eq:diffusion}\\
\Sigma_{\nu} & = & \min\left\{ \max\left[\alpha_{\nu}+\left(j_{\nu}+\chi_{\nu}\right)\frac{1}{2}\int f^{{\rm s}}_{\nu}d\mu,0\right],j_{\nu}\right\}, \nonumber \\
\alpha_{\nu} & = & \vec{\nabla}\cdot\left(\frac{-1}{3\left(j_{\nu}+\chi_{\nu}+\phi_{\nu}\right)}\vec{\nabla}f_{\nu}^{{\rm t}}\right),\nonumber 
\end{eqnarray}
is solved in 3D based on the angular integral $\frac{1}{2}\int f_{\nu}^{s}d\mu$
of the streaming neutrinos from the previous time step. In these equations,
$j_{\nu}$ is the spectral neutrino emissivity, $\chi_{\nu}$ the
neutrino absorptivity and $\phi_{\nu}$ includes isoenergetic scattering
in the mean free path (see e.g., \cite{bruenn1985}). The non-local
diffusion term $\alpha_{\nu}$ is evaluated by explicit finite differencing.
All other (local) variables are iterated to convergence by an implicit
Newton-Raphson scheme in each zone of the computational domain: Eq.
(\ref{eq:diffusion}) sets the net interaction rates \ensuremath{s_{\nu}}
 between matter and the radiation particles, which in turn determine the 
updates of the electron fraction $Y_{e}$ and the specific internal
energy $e$, 
\begin{eqnarray}
s_{\nu} & = & \frac{\partial f_{\nu}^{{\rm t}}}{c\partial t}+\Sigma_{\nu}-\left(j_{\nu}+\chi_{\nu}\right)\frac{1}{2}\int f_{\nu}^{{\rm s}}d\mu,\label{eq:trapped.update.matter}\\
\frac{\partial Y_{{\rm e}}}{c\partial t} & = & -\frac{m_{{\rm b}}}{\rho}\frac{4\pi c}{\left(hc\right)^{3}}\int\left(s_{\nu_{{\rm e}}}-s_{\bar{\nu}_{{\rm e}}}\right)E^{2}dE,\label{eq:trapped.update.ye}\\
\frac{\partial e}{c\partial t} & = & -\frac{m_{{\rm b}}}{\rho}\frac{4\pi c}{\left(hc\right)^{3}}\int\left(s_{\nu_{{\rm e}}}+s_{\bar{\nu}_{{\rm e}}}\right)E^{3}dE.\label{eq:trapped.update.e}
\end{eqnarray}
The rates $j_{\nu}$, $\chi_{\nu}$ and $\phi_{\nu}$ in Eq. (\ref{eq:diffusion})
can then be adjusted to the changes in $Y_{\text{e}}$ and $e$ until
convergence is achieved. In these equations $m_{\text{b}}$ is the
mass of a baryon, $c$ the light speed, and $h$ is Planck's constant.

Assuming a stationary-state solution and neglecting observer corrections
(see \cite{liebendoerfer2009} for a more detailed discussion of these assumptions)
we find that the integration of the net rate over the interior of
a sphere with radius $R$ delivers a useful approximation for the
neutrino number luminosity $4\pi R^{2}q_{\nu}$ at the surface of
the sphere. Hence, for the streaming neutrino flux we obtain
\begin{equation}
q_{\nu}=\frac{1}{4\pi R^{2}}\int_{0}^{R}\left(\frac{1}{2}\int\left[-\left(j_{\nu}+\chi_{\nu}\right)f_{\nu}^{{\rm s}}+\Sigma_{\nu}\right]d\mu\right)4\pi r^{2}dr.\label{eq:integratesrc}
\end{equation}

In the end, we are interested in the streaming neutrino density rather
than the flux. The spectral neutrino density is derived from the neutrino
flux by an analytically estimated flux factor $\mathcal{F}\left(E\right)$
\citep{liebendoerfer2009},
\[
\frac{1}{2}\int f_{\nu}^{s}d\mu=\frac{q_{\nu}\left(E\right)}{\mathcal{F}\left(E\right)}.
\]
This spectral neutrino density enters the diffusion Eq. (\ref{eq:diffusion})
of the next time step. Finally, the trapped neutrino fraction $Y_{\nu}$
and mean energy $Z_{\nu}$ are updated by integrating the resulting
spectral $f_{\nu}^{\text{t}}$ over energy. Moreover, the acceleration
of matter by trapped neutrino pressure differences is applied to the
velocity field,
\[
\frac{\partial v}{\partial t}=-\frac{1}{\rho}\frac{\partial}{\partial r}\left(\frac{\rho Z_{\nu}^{{\rm t}}}{3m_{{\rm b}}}\right).
\]

A mathematically rigorous analysis of the IDSA with all its limiting
cases is given in \cite{berninger2013}. The approximations are chosen
with care and always in comparison with the more comprehensive solution
of AGILE-BOLTZTRAN \citep{liebendoerfer2004}. We note that ELEPHANT includes
the crucial compressional heating of the trapped neutrino gas in Eq.
(\ref{eq:entropy.advection}), that is an $O(v/c)$ observer correction
in the original Boltzmann equation. Without this term, the trapped
neutrino gas would not even fulfill the hydrodynamic limit.

The computational domain of ELEPHANT is decomposed into $n$ rectangular
blocks of dimension $nx\cdot ny\cdot nz$ zones. Each block is assigned
to one MPI-task with distributed memory and enveloped by ghost zones
to support the largest stencil on the boundary zones. Moreover, the
block is sliced into sheets. Depending on a flag, the operations on
a sheet are either parallelized by OpenMP or sent to a GPU by OpenACC
statements. The OpenACC option currently is about a factor of two
faster than the OpenMP option, and is therefore the default option
used in this work.
Additional implementation details for ELEPHANT are given in \cite{kaeppeli2011}
and \cite{liebendoerfer2009}. 

\subsection{FLASH}
\label{flash}

The IDSA solver is also implemented in the publicly available FLASH code\footnote{\url{http://flash.uchicago.edu}} 
version~4 \citep{fryxell2000, dubey2008}. 
FLASH is a parallel, multi-dimensional hydrodynamic code based on block-structured adaptive mesh refinement (AMR). 
Our general setup is similar to that implemented by \cite{couch2014}, 
but we replace the radiation transfer solver with our multi-dimensional IDSA solver that is described 
in the previous section and in \cite{pan2016, pan2017b}.
In FLASH, the IDSA solver uses a similar strategy as in ELEPHANT, where we solve the diffusion source in multiple dimensions 
but keep the streaming component isotropic.

We use the third-order piecewise parabolic method (PPM, \citealt{colella1984}) for spatial reconstruction, and 
the HLLC Riemann solver and the Hybrid slope limiter for shock-capture. The nuclear equation of state unit 
in FLASH incorporates the finite temperature EOS routines from \cite{oconnor2010} 
and \cite{couch2014}\footnote{\url{http://www.stellarcollapse.org}}. 

Our current setup supports simulations in 1D spherical coordinates, in 2D cylindrical coordinates, 
and in 3D Cartesian coordinates. 
Self-gravity is calculated by the multipole solver from \cite{couch2013c} with $l_{\rm max}=16$ 
including the same effective GR potential correction as the other codes in this work. 

In addition to the GR potential correction, the main difference between the FLASH-IDSA in this paper 
and the IDSA implemented in \cite{pan2016} is that we have improved the ray sampling by performing 
a smoother grid-ray interpolation and have improved the IDSA ray resolution from $600$~zones to $1,000$~zones. 
Furthermore, in the current version of FLASH-IDSA, we disable the ray iteration in the streaming solver, 
but use a smaller neutrino time step ($dt_\nu=4 \times 10^{-7}$~s). 
By doing this, we obtain results that are more consistent with other SN groups 
(e.g., the s15 and s20 progenitor from \cite{woosley2007} failed to explode in 2D in Newtonian cases; 
\citealt{dolence2015, summa2016, suwa2016, pan2017a, oconnor2018}).

Graphics processing unit acceleration on the IDSA solver with OpenACC is also implemented in FLASH, 
but the parallelization strategy is different from that of ELEPHANT.
Instead of sending a sheet of data, we send the whole AMR block with one layer of guard cells to a GPU all at once.
While doing the calculation on GPUs, we transfer the next AMR block from host (central processing unit - CPU) to device (GPU) to avoid transfer overheads.
With a block size, $NXB=NYB=NZB=16$ and $NE=20$, we achieve an overall speedup of 2.3 with OpenACC, 
where $NXB$, $NYB$, $NZB$ are the block size in each dimension without guard cells and $NE$ is the number of energy bins.

Analysis and visualization of FLASH simulation data are provided by the {\tt yt} toolkit \citep{turk2011}.

\subsection{fGR1}
\label{fGR1}

The fGR1 code consists of three parts, where the evolution equations of the metric, hydrodynamics, and neutrino radiation are calculated.
Each of them is solved in an operator-splitting manner, but the system evolves self-consistently as a whole, satisfying the Hamiltonian and momentum constraints.
We briefly explain basic equations to be solved.
We note that a geometrized unit system is used in this subsection,
that is, the speed of light, the gravitational constant, and the Planck constant are set to unity: $c= G = h=1$.

The space-time metric is expressed in the standard (3+1) form: $ds^2=-\alpha^2dt^2+\gamma_{ij}(dx^i+\beta^idt)(dx^j+\beta^jdt),$ with $\alpha$ and $\beta^i$ being the lapse and shift, respectively.
The spatial metric $\gamma_{ij}$ and its extrinsic curvature $K_{ij}$ are evolved using the BSSN formulation \citep{shibata1995,baumgarte1999}  expressed in terms of the standard variables $\tilde\gamma_{ij}$, $W$, $K$, $\tilde A_{ij}$ and $\tilde\Gamma^{i}$, where $\tilde\gamma_{ij}=\psi^{-4}\gamma_{ij}$, with $\psi$ being the conformal factor, 
$W(=\psi^{-2})$  the inverse square of the conformal factor,
$K=\gamma^{ij}K_{ij}$  the trace of the extrinsic curvature, 
$\tilde A_{ij}=\psi^{-4}(K_{ij}-K\gamma_{ij}/3)$, 
and $\tilde\Gamma^{i}=-\partial_j \tilde\gamma^{ij}$ 
(see \cite{kuroda2012} for more details).

In the radiation hydrodynamics part, we solve the spectral neutrino transport equations, where the source terms are treated self-consistently following a procedure applying the M1 closure scheme \citep{shibata2011}.

The total stress-energy tensor $T^{\alpha\beta}_{\rm (total)}$ is expressed as 

\begin{equation} 
T_{\rm (total)}^{\alpha\beta} = T_{\rm (fluid)}^{\alpha\beta} +\int d\varepsilon \sum_{\nu\in\nu_e,\bar\nu_e,\nu_x}T_{(\nu,\varepsilon)}^{\alpha\beta},
\label{TotalSETensor}
\end{equation}

\noindent
where $T_{\rm (fluid)}^{\alpha\beta}$ and $T_{(\nu,\varepsilon)}^{\alpha\beta}$ are the stress-energy tensor of fluid and energy-dependent neutrino radiation 
field, respectively. We note that in the equation above, summation is taken over all species of neutrinos ($\nu_e,\bar\nu_e,\nu_x$) with $\nu_x$ representing heavy-lepton neutrinos (i.e., $\nu_{\mu}, \nu_{\tau}$ and their anti-particles). $\varepsilon$ represents the neutrino energy measured in the comoving frame with the fluid.
For simplicity, the neutrino flavor index is omitted below.

Introducing the radiation energy ($E_{(\varepsilon)}$), radiation flux ($F^{\mu}_{(\varepsilon)}$) and radiation pressure ($P^{\mu\nu}_{(\varepsilon)}$),
measured by an Eulerian observer or ($J_{(\varepsilon)}$, $H^{\mu}_{(\varepsilon)}$ and $L^{\mu\nu}_{(\varepsilon)}$) measured in a comoving frame, $T_{(\varepsilon)}^{\mu\nu}$  can be written in covariant form as

{\setlength\arraycolsep{2pt}
\begin{eqnarray}
T_{(\varepsilon)}^{\mu\nu}&=& E_{(\varepsilon)} n^\mu n^\nu+ F_{(\varepsilon)}^\mu n^\nu+
F_{(\varepsilon)}^\nu n^\mu +
P_{(\varepsilon)}^{\mu\nu},
\label{t_lab}\\
&=& J_{(\varepsilon)} u^\mu u^\nu+ H_{(\varepsilon)}^\mu u^\nu+
H_{(\varepsilon)}^\nu u^\mu +
L_{(\varepsilon)}^{\mu\nu}.
\label{t_com}
\end{eqnarray}}

In the above equations, $n^{\mu}=(1/\alpha,-\beta^k/\alpha)$ is a unit vector orthogonal to the space-like hypersurface and $u^\mu$ is the four velocity of the fluid.
In the truncated moment formalism \citep{thorne1981,shibata2011}, the radiation energy ($E_{(\varepsilon)}$) and radiation flux ($F^{\alpha}_{(\varepsilon)}$) are evolved in a conservative form and $P^{\mu\nu}_{(\varepsilon)}$ is determined by an analytic closure relation (e.g., Eq.~(\ref{neu_p})).
The evolution equations for $E_{(\varepsilon)}$ and $F^{\alpha}_{(\varepsilon)}$ are given by

\begin{eqnarray}
\partial_t \sqrt{\gamma}E_{(\varepsilon)}+\partial_i \sqrt{\gamma}(\alpha F_{(\varepsilon)}^i-\beta^i E_{(\varepsilon)})
+\sqrt{\gamma}\alpha \partial_\varepsilon \bigl(\varepsilon \tilde M^\mu_{(\varepsilon)} n_\mu\bigr)  =\nonumber \\
\sqrt{\gamma}(\alpha P^{ij}_{(\varepsilon)}K_{ij}-F_{(\varepsilon)}^i\partial_i \alpha-\alpha S_{(\varepsilon)}^\mu n_\mu),
\label{eq:rad1}
\end{eqnarray}

\noindent
and 

\begin{eqnarray}
\partial_t \sqrt{\gamma}{F_{(\varepsilon)}}_i+\partial_j \sqrt{\gamma}(\alpha 
{P_{(\varepsilon)}}_i^j-\beta^j {F_{(\varepsilon)}}_i)
-\sqrt{\gamma}\alpha \partial_\varepsilon\bigl(\varepsilon \tilde M^\mu_{(\varepsilon)} \gamma_{i\mu}\bigr)=\nonumber \\
\sqrt{\gamma}[-E_{(\varepsilon)}\partial_i\alpha +{F_{(\varepsilon)}}_j\partial_i \beta^j+(\alpha/2) 
P_{(\varepsilon)}^{jk}\partial_i \gamma_{jk}+\alpha S^\mu_{(\varepsilon)} \gamma_{i\mu}],
\label{eq:rad2}
\end{eqnarray}

\noindent
respectively.

Here, $\gamma$ is the determinant of the three metric, $\gamma\equiv{\rm det}(\gamma_{ij})$, $S^{\mu}_{(\varepsilon)}$ is the source term for neutrino matter interactions, and $\tilde M^\mu_{(\varepsilon)}$ is defined by $\tilde M^\mu_{(\varepsilon)}\equiv M^{\mu\alpha\beta}_{(\varepsilon)}\nabla_\beta u_\alpha$, where $M^{\mu\alpha\beta}_{(\varepsilon)}$ denotes the third rank moment of neutrino distribution function (see \citet{shibata2011} for the explicit expression).

By adopting the M1 closure scheme, the radiation pressure can be expressed as

\begin{eqnarray}
P_{(\varepsilon)}^{ij}=\frac{3\chi_{(\varepsilon)}-1}{2}P^{ij}_{\rm thin(\varepsilon) }
+\frac{3(1-\chi_{(\varepsilon)})}{2}P^{ij}_{\rm thick(\varepsilon)},
\label{neu_p}
\end{eqnarray}

\noindent
where $\chi_{(\varepsilon)}$ represents the variable Eddington factor, and $P^{ij}_{\rm thin(\varepsilon)}$ and $P^{ij}_{\rm thick(\varepsilon)}$ correspond 
to the radiation pressure in the optically thin and thick limit, respectively. They are written in terms of $J_{(\varepsilon)}$ and $H^\mu_{(\varepsilon)}$ \citep{shibata2011}. Following \citet{minerbo1978, cernohorsky1994}, and \citet{obergaulinger2011}, we take the variable Eddington factor $\chi_{(\varepsilon)}$ as

\begin{eqnarray}
\label{eq:Closure_Chi}
\chi_{(\varepsilon)}&=&\frac{5+6\bar{F}^2_{(\varepsilon)}-2\bar{F}^3_{(\varepsilon)}+6\bar{F}^4_{(\varepsilon)}}{15},
\end{eqnarray}

\noindent
where,

\begin{eqnarray}
\label{eq:F^2}
\bar{F}^2_{(\varepsilon)}&\equiv&\frac{h_{\mu\nu}H^\mu_{(\varepsilon)} H^\nu_{(\varepsilon)}}{J^2_{(\varepsilon)}}.
\end{eqnarray}

In Eq.~(\ref{eq:F^2}), $h_{\mu\nu}\equiv g_{\mu\nu}+u_\mu u_\nu$ is the projection operator. By the definition of $\bar{F}_{(\varepsilon)}$ in Eq.~(\ref{eq:F^2}), we can appropriately reproduce several important neutrino behaviors; for example, neutrino trapping in the rapidly collapsing opaque core \citep{kuroda2016}.
We simultaneously solve Eqs.(\ref{eq:Closure_Chi}-\ref{eq:F^2}) using the Newton method to find a converged solution of $\chi_{(\varepsilon)}$.

The hydrodynamic equations are written in a conservative form as

\begin{eqnarray}
\label{eq:GRmass}
\partial_t \rho_{\ast}&+&\partial_i(\rho_\ast v^i)=0,\\
\label{eq:GRmomentum}
\partial_t \sqrt{\gamma} S_i&+&\partial_j \sqrt{\gamma}( S_i v^j+\alpha P\delta_i^j)=\nonumber \\
&&-\sqrt{\gamma}\biggl[ S_0\partial_i \alpha- S_k\partial_i \beta^k-2\alpha S_k^k\partial_i \phi \nonumber \\
&&+\alpha e^{-4\phi} ({S}_{jk}-P \gamma_{jk}) \partial_i 
\tilde{\gamma}^{jk}/2+\alpha \int d\varepsilon S_{(\varepsilon)}^\mu \gamma_{i\mu} \biggr],\nonumber\\ \\
\label{eq:GRenergy}
\partial_t \sqrt{\gamma} \tau&+&\partial_i \sqrt{\gamma}(\tau v^i+P(v^i+\beta^i))=\nonumber \\
&&\sqrt{\gamma}\biggl[ \alpha K S_k^k /3+\alpha e^{-4\phi} ({S}_{ij}-P \gamma_{ij})\tilde{A^{ij}}  \nonumber\\
&&- S_iD^i\alpha+\alpha \int d\varepsilon S_{(\varepsilon)}^\mu n_\mu \biggr],\\
\label{eq:GRYe}
\partial_t (\rho_\ast Y_e)&+&\partial_i (\rho_\ast Y_e v^i)=\sqrt{\gamma}\alpha m_{\rm u}\int \frac{d\varepsilon}{\varepsilon}
(S_{(\nu_e,\varepsilon)}^\mu-S_{(\bar\nu_e,\varepsilon)}^\mu) u_\mu,
\end{eqnarray}

\noindent
where $\rho_\ast=\rho \sqrt{\gamma}W$, $S_i=\rho hW u_i $, $S_{ij}=\rho h u_i u_j+P\gamma_{ij}$, $S_k^k=\gamma^{ij}S_{ij}$, $S_0=\rho h W^2-P$, and $\phi={\rm log}(\gamma)/12$. The magnitude $\rho$ is the rest mass density, $W$ is the Lorentz factor, $h= 1+e+P/\rho$ is the specific enthalpy, $v^i=u^i/u^t$, $\tau= S_0-\rho W$, $Y_e\equiv n_e/n_b$ is the electron fraction ($n_X$ is the number density of $X$), $e$ and $P$ are the specific internal energy and pressure of matter, respectively, and $m_{\rm u}$ is the atomic mass unit. $P(\rho,s,Y_e)$ and $e(\rho,s,Y_e)$ are given by an EOS with $s$ denoting the entropy per baryon. On the right-hand side of Eq.~(\ref{eq:GRenergy}), $D^i$ represents the covariant derivative with respect to the three metric $\gamma_{ij}$.

fGR1 employs a pure MPI parallelization scheme and enforces a refluxing procedure at the refinement boundary to ensure conservation laws.

\subsection{SPHYNX}
\label{sphynx}

The application of SPH to simulate CCSNe is not new. As early as in 1992, SPH was used to perform 2D simulations \citep{herant1992}, highlighting the roles of the convective overturn and hydrodynamical instabilities in aiding the explosion. In 2002, SPH was used to perform the very first 3D simulations of CCSNe \citep{fryer2002} and in 2004 again, this time including rotation \citep{fryer2004}. These works were undoubtedly remarkable landmarks and the use of SPH helped to achieve this outcome
for three main reasons: firstly, SPH is intrinsically devised for 3D simulations without boundaries \citep{lucy1977,gingold1977}; secondly, it has an adaptive spatial resolution that, in general terms, follows mass density, allowing a highly resolved proto-neutron star without presenting an unaffordable computational burden; and finally, it conserves energy and momentum (both linear and angular) by construction, which makes it very suitable for simulating rotating models, where momentum transfer is critical.

In our previous work \citep{perego2016} we proved the capabilities of using SPH to simulate CCSNe employing a novel spectral leakage treatment (ASL). To our knowledge, this was the first time since the early works of Herant et al. and Fryer et al. that SPH had been used to simulate this scenario, and the very first time that it was done using a spectral treatment of the neutrino component within the SPH framework. In this work, we extend the neutrino treatment, coupling the IDSA with our SPH code, SPHYNX\footnote{Available at \url{http://astro.physik.unibas.ch/sphynx}}.

SPHYNX includes some of the latest upgrades that have enabled a remarkable advancement in this numerical technique during recent years. Among them are: calculating gradients using an integral approach to derivatives (IAD), which has been proven to provide more accurate derivatives \citep{garcia2012,cabezon2012,rosswog2015,valdarnini2016}; using a high-order, paring-resistant family of kernels ($sinc$-kernels), which provides higher flexibility \citep{cabezon2008}; and implementing novel generalized volume elements, which prevent the emergence of the tensile instability and facilitate the development of hydrodynamical instabilities \citep{cabezon2017}.

SPHYNX solves the Euler equations, derived from a variational principle (see \cite{rosswog2009} and references therein). It also evaluates 3D gravity with a multipolar approximation calculated via a hierarchical tree structure that is based on the Barnes-Hut algorithm \citep{hernquist1989}. We have additionally included the same effective GR potential correction \citep{marek2006} as in ELEPHANT and FLASH to replace the monopolar term of the gravitational force. Neutrinos are handled with the IDSA treatment \citep{liebendoerfer2009}. Neutrino-electron scattering effects during collapse are taken into account by evolving the electron fraction and entropy with the parametrized deleptonization scheme of \cite{liebendoerfer2005}. 

To evolve the system with SPHYNX we solve the hydrodynamical equations in Lagrangian form:

\begin{eqnarray}
 \frac{ {\rm d}\rho}{\rm dt} & = & -\rho \nabla \cdot \mathbf{v}, \label{sphdens}\\
 \frac{ {\rm d}\mathbf{v}}{\rm dt} & = & \frac{-\nabla P_{\rm tot}}{\rho}+\mathbf{g}, \label{sphmom}\\
 \frac{ {\rm d}u}{\rm dt} & = & \frac{-P_{\rm gas}}{\rho}\nabla \cdot \mathbf{v}+\dot{e}_{\nu,I}, \label{sphene}\\
 \frac{ {\rm d}Z_{\nu_k}}{\rm dt} & = & \frac{-P_{\nu_k}}{\rho}\nabla \cdot \mathbf{v}+\dot{Z}_{\nu_k,I}, \label{sphznu}\\
 \frac{ {\rm d}Y_{\nu_k}}{\rm dt} & = & \dot{Y}_{\nu_k,I}, \label{sphynu}\\
 \frac{ {\rm d}Y_e}{\rm dt} & = & \dot{Y}_{e,I}, \label{sphye}
\end{eqnarray}

\noindent
where $\mathbf{v}$ is the velocity vector, $P_{\rm tot} = P_{\rm gas} + P_{\nu}$, and

\begin{equation}
 P_{\nu}=\sum_k P_{\nu_k} = \rho \, \sum_k Z_{\nu_k} /\left( 3 m_b \right).
\label{pnu1}
\end{equation}

\noindent
The index $k$ runs over all (anti)neutrino species. As a result, Eq.~(\ref{sphmom}) includes the extra contribution of the neutrino radiation field, and therewith, the stress provided by the trapped neutrinos is directly taken into account. 

From the position of the SPH particles and the equation of state, we compute the local density ($\rho$), the gradient of the pressure ($P$), the gravitational acceleration ($\mathbf{g}$), and the internal energy ($u$). Moreover, each particle carries information regarding the electron fraction ($Y_e$), the neutrino abundances and energies related to the neutrino trapped component, ($Y_{\nu}, Z_{\nu}$), for all neutrino flavors. The IDSA ultimately provides the rates of change for these quantities, ($\dot{Y}_{e,I}$, $\dot{Y}_{\nu,I}$, $\dot{Z}_{\nu,I}$), and for the internal energy, ($\dot{e}_{\nu,I}$), as denoted by the subscript $I$. 

The abundances of electrons and neutrinos are evolved explicitly from Eqs.~(\ref{sphynu}) and (\ref{sphye}). It is worth noting that, as can be seen in Eq.~(\ref{sphznu}), we implement an additional energy-like equation to evolve the energy of the trapped component of the neutrinos. In this way, we can take into account both the contribution due to the compression of the neutrino field and the rate of change provided by the IDSA, in a consistent and robust way. The details on the implementation of Eq.~(\ref{sphznu}) can be found in Appendix~C of \cite{perego2016}.

The IDSA is primarily a local treatment of the neutrinos. Nevertheless, it requires the spectral optical depth and the distribution function of streaming neutrinos, which are non-local quantities. Additionally, a diffusion equation for the distribution function of trapped neutrinos has to be solved. For the first two, we employ the same approach as in ELEPHANT and FLASH: we assume that the streaming component is isotropic and use spherically averaged mean free paths to integrate the optical depth. Regarding the solution of a diffusion equation, which implies second derivatives, we use the approach of \cite{brookshaw1985}, which is less sensitive to particle distribution and has been widely used in SPH simulations with radiative transfer (see e.g., \cite{whitehouse2004} and \cite{jubelgas2004} for a reformulation):

\begin{equation}
 \alpha_i=\frac{{\rm d}f^t_i}{{\rm d}t}=\frac{c}{3}\sum_j\frac{m_j}{\rho_i\rho_j}\frac{\left(\lambda_i+\lambda_j\right)\left(f^t_j-f^t_i\right)}{\left|\mathbf{r}_{ij}\right|^2}\mathbf{r}_{ij}\cdot\nabla_iW_{ij},
\label{alpha_sph}
\end{equation}

\noindent
where $f^t$ is the distribution function of the trapped component of neutrinos, $\lambda$ is the mean free path, while subscripts $i$ and $j$ refer to a particle and its corresponding neighbors, respectively. In this way, we obtain $\alpha_i$,  the diffusion source for each SPH particle, as defined in Eq.~7 from \cite{liebendoerfer2009}.

SPHYNX uses a hybrid MPI+OpenMP parallelization scheme. Its capabilities, implementation, and results in several tests can be found in \cite{cabezon2017}.

\subsection{Implementation details}
\label{implementation}

\begin{table*}[h]
\begin{center}
\begin{tabular}{|l|l|l|l|l|l|}
\hline
\multicolumn{1}{|c}{Code}&\multicolumn{1}{|c}{$\nu-$Transport method}&\multicolumn{1}{|c}{NES}&\multicolumn{1}{|c}{Gravity}&\multicolumn{1}{|c}{Rotation}&\multicolumn{1}{|c|}{Progenitor}\\\hline\hline
E \dots ELEPHANT &I \dots IDSA &$(-)$ \dots w/o NES&$(-)$\dots Newt.&$(-)$ \dots No&95 \dots s15s7b2(W95)\\ 
F \dots FLASH &P \dots IDSA$+$PD&S \dots w NES&G \dots GR&R \dots Rapid &07 \dots s15(W07)\\ 
S \dots SPHYNX &M \dots M1&& & &\\
G \dots fGR1 & && & &\\
X \dots All codes &&&&&\\ \cline{1-2}
\multicolumn{2}{|l|}{B  \dots  AGILE-BOLTZTRAN}&& & &\\  \hline
\end{tabular}
\end{center}
\caption{Summary of abbreviations used for the naming of runs. A dash means that no letter is added.}
\label{table:naming}
\end{table*}

In this section we discuss the setup of the simulations in the framework of every code. We also point to the similarities and differences regarding the physics implementation, domain discretization, and spatial resolution, among the different codes.

\subsubsection{Initial conditions and equation of state}
In order to link to both legacy models and newer ones, we
start from 1D progenitor models s15s7b2 and s15 from \cite{woosley1995}
and \cite{woosley2007}, respectively. All codes begin with the same initial conditions: the onset of
the collapse of a massive star with $15$~\msun at zero age main sequence.
We map the density, temperature,
electron fraction, and radial velocity profiles to the 3D computational
domain. This process is adapted to each code, but the resulting initial 3D radial profiles are very similar in all codes.

 In this study all codes use the Lattimer-Swesty EOS \citep{lattimer1991} with incompressibility K = 220 MeV (LS220).
Although recent studies have shown that the LS220 EOS does not fulfill some theoretical 
and experimental nuclear constraints \citep{kruger2013}, this is one of the most popular EOS in the SN community 
and has been widely used in many CCSN simulations. Moreover, the study of the effect of different EOSs in CCSN simulations is not the objective of this work.

\subsubsection{Additional physics, discretization, and spatial resolution}
\label{additionalphysics}
All three Newtonian codes (ELEPHANT, FLASH, and SPHYNX) have a 3D gravity solver that has a corrected monopolar term
to use the effective GR potential of \cite{marek2006}. Additionally, depending on the run, they can use the parametrized deleptonization (PD) scheme during the collapse phase 
to mimic the effects from neutrino-electron scattering (NES) \citep{liebendoerfer2005}. In this case, during the
collapse phase, the IDSA solver is only used to update 
the background neutrino fraction and energies. After core bounce, the PD scheme is always turned off and the IDSA takes over. In contrast to this approximation, fGR1 calculates true NES rates and includes this process in its two-moment M1 neutrino transport scheme.  

An effective GR potential correction (Case~A in \citealt{marek2006}) is used in all Newtonian codes to replace the monopole term in their respective gravity solvers, 

\begin{equation}
\Phi_{GR}(r) = \Phi_{NW}(r) - \Phi_{NW,~l=0}(r) + \Phi_{eff.~GR,~l=0}(r)\,,
\end{equation}

\noindent
where $\Phi_{GR}$ and $\Phi_{NW}$ are the GR and Newtonian gravitational potentials, respectively, $\Phi_{eff.~GR}$ stands for the effective GR correction, and $l=0$ denotes the monopole term.

The modeling of $\mu$/$\tau$ neutrinos in the IDSA runs is handled by an inexpensive gray leakage scheme largely based on the scheme presented in the appendix of \cite{rosswog2003}. Nevertheless, it is worth mentioning that electron neutrinos and heavy neutrinos are in fact coupled, for example by direct annihilation/creation processes, but because these cross-flavor processes are not included in our present IDSA version, electron neutrinos are decoupled from heavy neutrinos in our simulations. More generally however, caution with respect to the results of the leakage scheme is in place, especially for the longer-term evolution of the post-bounce phase. But this is not the focus of this comparison and so we feel safe to evolve them using different numerical algorithms. In our scheme, the energy loss caused by the emission of heavy flavor neutrinos is computed as smooth interpolation between diffusion and production rates. The former are 
valid in optically thick regions, the latter in free-streaming conditions. Neutrino elastic scattering off nucleons \citep{bruenn1985} is employed as an opacity source to compute the relevant energy-averaged optical depth along the radial path. Electron-positron annihilation rates integrated over the relevant phase space are considered as the dominant production channel and are implemented according to \cite{itoh1996}. The relevant diffusion rate is calibrated against post-bounce results obtained with detailed Boltzmann neutrino transport (AGILE-BOLTZTRAN code) for a 15~\msun progenitor (s15s7b2). 

Regarding the energy resolution, all four codes use logarithmically spaced energy bins. The three IDSA implementations cover 3 to 300~MeV with 20 energy bins,
while fGR1 covers 1 to 300~MeV with 12 energy bins. The difference in energy resolution is basically due to the more detailed neutrino transport of fGR1. Being more costly than IDSA, it partially compensates the computational load with less energy bins. Neutrino energies in fGR1 are measured in the comoving frame.

For this comparison we use a resolution of
the equidistant mesh in ELEPHANT of $\unit[1]{km}$. The 3D computational domain
extends to $(\unit[300]{km})^3$ and is embedded in a larger spherically
symmetric computational domain that is evolved by the spherically
symmetric implicitly finite differenced code AGILE-IDSA. AGILE-IDSA
provides the boundary conditions for the 3D domain at $\unit[300]{km}$
radius and is used as a limiter for the entropy at densities larger than $\unit[10^{14}]{g~cm^{-3}}$,
at the very center of the PNS. The latter measure was
necessary to prevent an unphysical updrift of the central entropy
due to the numerical dispersion of the steep entropy gradient between
the unshocked innermost part and the shock-heated outer part of the
PNS in the 3D hydrodynamics code. 
  
In FLASH, we use 3D Cartesian grids for this study. 
The center of the progenitor star is located at the origin of the simulation box, 
which includes $\pm 5,000$~km in each spatial direction.
The central region $r \leq 32$~km has the smallest zone size of $0.488$~km. 
Additional AMR decrements based on the distance to the origin are imposed to save computational time, 
giving an effective angular resolution of $0.9^\circ - 1.7^\circ$.

All SPHYNX simulations presented here use 200000 SPH particles and cover the inner
$\sim1.8$~\msun of the progenitor star, which includes 
the whole iron core and part of the silicon layer. As it is characteristic for SPH codes, the spatial resolution automatically follows the density, reaching $0.477$~km in the PNS. Outer layers are taken into account via an external pressure that is applied to the whole domain. The value of this pressure is taken from the 1D initial profile and it affects only
the most external SPH particles.

In fGR1, the 3D computational domain is a cubic box
of 15,000 km in width where nested boxes with nine refinement
levels are embedded in Cartesian coordinates. 
To save computational time however, the initial refinement level is set to five which is increased when the maximum density reaches $5\times10^{10}$, $10^{12}$, $10^{13}$, and $3\times10^{13}$ g cm$^{-3}$ during the collapse phase.
Each box contains $64^3$ cells and the
minimum grid size near the origin is $\Delta x=458$~m.
The PNS core surface ($\sim10$~km) and stalled shock ($\sim110$-220~km)
are resolved by $\Delta x=458$~m and $\Delta x=7.3$~km, respectively.

When rotation is included in the calculation, we implement a shellular profile, following the setup of \cite{yokozawa2015}. We discuss further details of the rotating models in Sect.~\ref{rotation}.

\section {Data summary, definition of acronyms of runs, and abbreviation table}
\label{sec:datasummary}
CCSN simulations rely on many different aspects. For this reason, making a detailed comparison of all of them is out of the scope of this work. Nevertheless, we focus on three main pillars to compare the outcomes of our simulations: the neutrino-transport method, the gravity treatment, and the inclusion of rotation. Our runs include different combinations of these three aspects for each of the hydrodynamical codes of Sect.~\ref{sec:methods}. 
Here we give an overview of the performed runs as well as an explanation of their names and abbreviations for ease of reference in the different comparisons.
\begin{table*}[t!]
\begin{center}
\begin{tabular}{|c|c|c|c|c|l|}
\hline
Code&$\nu-$Transport &Gravity&Rotation&Progenitor model&\multicolumn{1}{c|}{Name of run}\\ \hline \hline 
\multirow{8}{*}{ELEPHANT}&\multirow{3}{*}{IDSA}&Newt.&\multirow{2}{*}{No}&\multirow{2}{*}{s15s7b2 (W95)}& E95-I\\ \cline{3-3}
   &   & GR      &     &            &E95-IG   \\ \cline{3-5}
   &   & Newt.   & Yes & s15 (W07)   &E07-IR   \\ \cline{2-5}

   &\multirow{4}{*}{IDSA+PD}&Newt.   & \multirow{2}{*}{No}   &  \multirow{2}{*}{s15s7b2 (W95)} & E95-P\\ \cline{3-3}
   &   &  GR &   &   &E95-PG \\\cline{3-5}
   &   &   \multirow{2}{*}{Newt.}  & Yes &  \multirow{2}{*}{s15 (W07)}    & E07-PR \\ \cline{4-4}
   &   &                 &  No   &      & E07-P \\ \hline\hline 
\multirow{8}{*}{FLASH}&\multirow{3}{*}{IDSA}&Newt.&\multirow{2}{*}{No}&\multirow{2}{*}{s15s7b2 (W95)}& F95-I\\ \cline{3-3}
   &   &GR  &    &    &F95-IG \\ \cline{3-5}
   &   & Newt.   & Yes & s15 (W07)   &F07-IR\\ \cline{2-5}

   &\multirow{4}{*}{IDSA+PD}&Newt.   & \multirow{2}{*}{No}   &  \multirow{2}{*}{s15s7b2 (W95)} & F95-P\\ \cline{3-3}
   &   &  GR &   &   &F95-PG \\\cline{3-5}
   &   &   \multirow{2}{*}{Newt.}  & Yes &  \multirow{2}{*}{s15 (W07)}    & F07-PR \\ \cline{4-4}
   &   &                 &  No   &      & F07-P \\ \hline\hline 
\multirow{8}{*}{SPHYNX}&\multirow{3}{*}{IDSA}&Newt.&\multirow{2}{*}{No}&\multirow{2}{*}{s15s7b2 (W95)}& S95-I\\ \cline{3-3}
   &   &GR  &    &    &S95-IG \\ \cline{3-5}
   &   & Newt.   & Yes&s15 (W07)&S07-IR\\ \cline{2-5}
  
   &\multirow{4}{*}{IDSA+PD}&Newt.   & \multirow{2}{*}{No}   &  \multirow{2}{*}{s15s7b2 (W95)} & S95-P\\ \cline{3-3}
   &   &  GR &   &   &S95-PG \\\cline{3-5}
   &   &   \multirow{2}{*}{Newt.}  & Yes &  \multirow{2}{*}{s15 (W07)}    & S07-PR \\ \cline{4-4}
   &   &                 &  No   &      & S07-P \\ \hline\hline 
\multirow{2}{*}{fGR1} &  M1$-$NES & \multirow{2}{*}{GR}  & \multirow{2}{*}{No} & \multirow{2}{*}{s15s7b2 (W95)}&G95-MG \\\cline{2-2}
       & M1+NES & & & & G95-MSG \\ \hline\hline 
\multirow{4}{*}{AGILE-BOLTZTRAN}&\multirow{2}{*}{Boltzmann$-$NES}& Newt.& \multirow{4}{*}{No}&\multirow{2}{*}{s15s7b2 (W95)}&B95\\ \cline{3-3}
   &   & GR  &   &      & B95-G\\ \cline{2-3} \cline{5-5}
   &\multirow{2}{*}{Boltzmann+NES}& Newt.&  &\multirow{2}{*}{s15s7b2 (W95)}&B95-S\\ \cline{3-3}
   &   & GR  &   &      & B95-SG\\ \cline{3-3} \cline{5-5}\hline 
\end{tabular}
\end{center}
\caption{Overview of the combinations of hydrodynamic codes and physics included in our runs, with their corresponding names.}
\label{table:overview}
\end{table*} 

The name of a run gives information on the aforementioned pillars and codes. To easily distinguish which code and progenitor has been used with which settings, the name of a run is twofold: the first part indicates the code and progenitor, and the second part shows the chosen setting of gravity, rotation, and (with the exception of AGILE-BOLTZTRAN) the $\nu$-transport method. The structure of the resulting abbreviations of the performed simulations is given by 

\begin{equation}
\begin{split} 
 &(\textrm{Name of run})=\\
 &[(\mbox{Code})(\textrm{Progenitor})]-[(\nu-\textrm{Transport})(\mbox{Gravity})(\mbox{Rotation})],
\end{split}
\label{eq_name1}
\end{equation}

The bracket on the RHS depends on the specific setting of a run and is set as given in Table~\ref{table:naming}. Below, an example is provided to illustrate the naming generated by Eq.~(\ref{eq_name1}). It describes an ELEPHANT run using the s15s7b2 progenitor from \cite{woosley1995} with IDSA as neutrino transport, Newtonian gravity and no rotation. The dash thereby indicates that no letter is added to the run name.

\begin{align*}
\textrm{E}95-\textrm{I}=[(\textrm{E})(95)]-[(\textrm{I})(-)(-)]
\end{align*}

Table~\ref{table:naming} shows an overview of abbreviations. The four 3D hydrodynamical codes (ELEPHANT, FLASH, SPHYNX and fGR1) are abbreviated by using their first capital letter. An X as code name is used as a wildcard, stating \textit{all codes} with a specific configuration. We use four neutrino treatments: IDSA, IDSA with parametrized deleptonization (PD), and the M1 scheme with and without NES.

General relativity and fast rotation are only considered by their first letter in the naming code, if applied. As progenitors, we use two 15~\msun models (s15s7b2 and s15) from \cite{woosley1995} and \cite{woosley2007}, respectively, referring to them by the last two digits of their publication year.
Table~\ref{table:overview} gives a complete overview of the runs performed and shall be a guide to the many run names used in this work.

\section{Comparison}
\label{sec:comparison}
We present here the results of our simulations with the codes introduced in Sect.~\ref{sec:methods} and compare them to each other. In order to ease the reading of the plots, we kept the same color for the same code among all plots independently of the implemented physics or used progenitor, unless stated otherwise. Namely, ELEPHANT results are always presented in blue, FLASH in green, SPHYNX in red, and fGR1 in magenta. To show the neutrino treatment, progenitor, and gravity evaluation we follow the naming formalism introduced in Sect.~\ref{sec:datasummary}.

\subsection{Neutrino transport}
\label{transport}

\begin{figure}[htbp]
\begin{center}
\includegraphics[width=\linewidth]{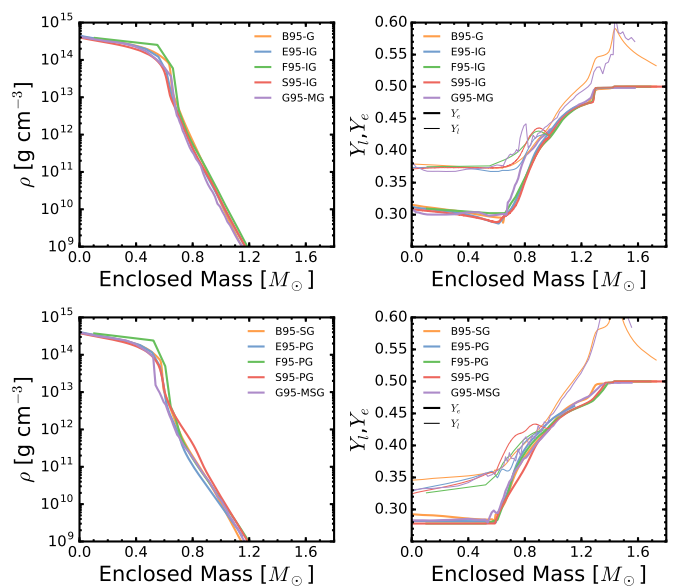}
  \caption{The rest mass density (left column) and electron/lepton fraction profiles as a function of the enclosed baryon mass for various models at bounce.
  Upper and lower rows are without and with taking into account the NES effect, respectively.
\label{KTf2}
}
\end{center}
\end{figure}

The nature of the following comparison is twofold: on the one hand,
we compare different approximations to neutrino transport (spherically
symmetric Boltzmann solver \citep{liebendoerfer2004} versus M1 in
general relativity \citep{kuroda2012} versus the isotropic diffusion
source approximation (IDSA, \citealt{liebendoerfer2009}) in FLASH,
SPHYNX and ELEPHANT). On the other hand, we compare three different
implementations of the same IDSA equations in the three latter codes.
While we succeed to a large extent in using the same progenitor star,
equation of state, and reaction rates as input physics, the implementations
of the hydrodynamics equations in the three codes differ significantly.
It therefore remains difficult to clearly attribute differences in the
results to either the neutrino transport or the hydrodynamics part.

The differences between the results of simulations that implement
the same set of equations should converge to zero if the resolution
of the simulation is increased. Global astrophysical simulations
however, span a dynamical range of many orders of magnitude. Thus,
high resolution is excessively expensive. Often, it is neither possible
nor reasonable to obtain sufficient computational resources to perform
simulations that are ``converged'' throughout the computational
domain in a mathematical sense. It is much more important to limit
the deviations from an exact solution in under-resolved regimes by
forcing the simulation to obey fundamental conservation laws and space-
or time-averaged interaction rates, even if the spatial or temporal
resolution is not sufficient throughout thus allowing  a mathematically converged
solution to be obtained.

Additionally, one cannot even be sure that a mathematically converged
solution exists. The SN scenario involves large-scale convection
of eddies between the surface of the proto-neutron star and the
shock front. Small-scale
turbulence connected to magnetic fields might also play a significant
role. Therefore, it is likely that there always remain regions in a simulation where
the evolution of small-scale features is by principle unpredictable, although
it remains macroscopically constrained by conservation laws and average
interaction rates. In
this sense it might be in vain to strive for mathematical convergence
with respect to all quantities in a simulation. Nonetheless, based
on the observation of SNe one can expect that some
overall physics features of the explosion remain robust with respect
to perturbations: for example, the occurrence of the collapse of the stellar core, the
occurrence of bounce at nuclear density, the formation of an initially
rather spherical standing accretion shock, the occurrence of fluid
instabilities, the expansion of the shock, the development of asymmetries,
the explosion, and important features of the nucleosynthesis.

Hence, we believe that it is justifiable that astrophysics codes show
slightly different results even if they are ``solving'' the same
set of equations. They should, however, agree overall on all features
that are observable and common to the majority of observed SNe.

\subsubsection{Stationary state approximation in the IDSA}

\begin{figure*}
\begin{center}
\includegraphics[width=\linewidth]{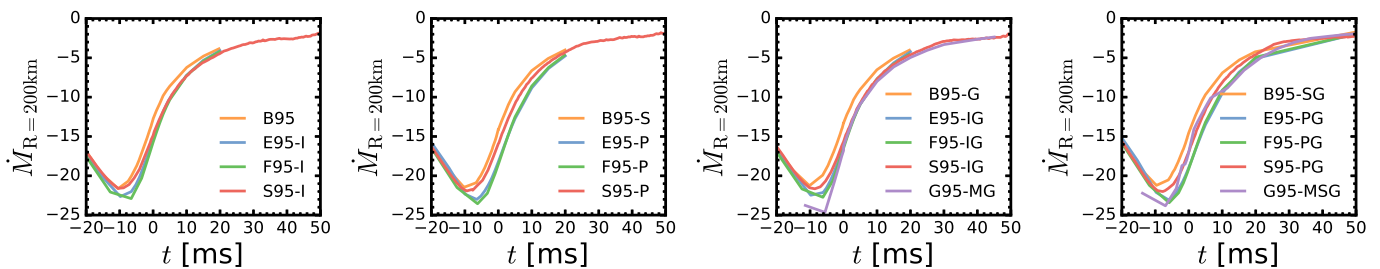}
\caption{Time evolution of mass accretion rates measured at $R=200$~km. Different colors represent simulations with different codes. Each column shows a set of simulations (see Table~\ref{table:overview}). 
\label{fig:mdot}}
\end{center}
\end{figure*}

Matter in the gravitational collapse of a massive star has a low entropy
around $\unit[1]{k_{B}}$ per baryon. Neutron-rich heavy nuclei dominate the
scattering opacity. Due to the low ratio of neutrino energy versus
the mass of a representative heavy nucleus, the scattering of trapped
neutrinos is nearly elastic. A downward shift of the mean neutrino
energies can still occur via inelastic neutrino-electron scattering,
where both scattering partners have more similar energies. Scattering
off electrons happens at a much lower rate than scattering off nuclei, but
for trapped neutrinos it is efficient enough to allow more neutrinos
to escape from the collapsing core due to their down-scattered energy
(see e.g., \citealt{cooperstein1988, bruenn1989, bethe1990} for a more
detailed explanation). This process cannot be modeled consistently
with the stationary state assumption of the IDSA. The IDSA neglects
the time that neutrinos need to propagate from a location A of creation
to another location B of further interaction \citep{liebendoerfer2009}. 
In a more recent implementation of the IDSA with time-dependent
propagation of streaming neutrinos it was possible to include neutrino-electron
scattering \citep{takiwaki2014}. For similar reasons, the original
implementation of the IDSA cannot adequately take into consideration the propagation
of heavy neutrinos through the stellar core. 

We solve these problems pragmatically,  using a simple and (currently) popular method to parameterize the effects of neutrino-electron
scattering during the collapse phase \citep{liebendoerfer2005} and a gray leakage scheme for the
emission of mu-neutrinos and tau-neutrinos (see Sect. 2.5.2). Such effective treatment, 
once calibrated against more detailed transport schemes, can capture the most relevant cooling 
effect provided by the emission of heavy flavor neutrinos and the related evolution of the PNS radius 
with reasonable accuracy. Nevertheless, we stress that the investigation of more subtle effects related
for example to the detailed treatment of neutrino-matter opacities would require a more sophisticated
treatment for mu and tau neutrinos as well.

After core-bounce, the
nuclei in the trapped regime become shock-dissociated. Then, the emission
and absorption of neutrinos on free nucleons dominate possible
effects of neutrino-electron scattering such that the approximations
of the IDSA should become adequate to model the post-bounce evolution,
whether neutrino-electron scattering is included or not.

The results of the collapse phase with and without parameterized deleptonization
are compared in Sect. 4.2. In this section we focus on the post-bounce
evolution where both transport methods, the IDSA, and full transport
can be compared.

\subsubsection{Comparison of data with respect to neutrino transport}

In Fig.~\ref{KTf2} the rest mass density and composition is shown at the time of bounce.
The figure documents that the models start into the post-bounce phase with reasonably small
initial deviations. From this point on, it is only the accretion of matter
from the outer layers that controls the further evolution of the post-bounce phase. The
mass accretion rate as a function of time is shown in Fig.~\ref{fig:mdot} at a radius of $200$~km.
The accretion rate acts like a boundary condition for the evolution
of the early post-bounce phase.
In the Newtonian runs, the two grid-based codes show a $\sim10\%$ larger accretion than the
two Lagrangian codes. However, this grouping breaks up in the general relativistic runs: SPHYNX,  in GR,
shows a slightly higher accretion rate than in the Newtonian case, while fGR1 seems to struggle
with resolution in the early collapse phase.
We note that the accretion rate evolution in SPHYNX changes slightly at 200~km when comparing Newtonian and GR-corrected simulations. This is related with the gravitational softening that is used to prevent divergences when two particles get too close. When we replace the monopolar term with the GR potential this softening is partially removed, effectively making the gravitational potential slightly stronger.

\begin{figure*}
\begin{center}
\includegraphics[width=\linewidth]{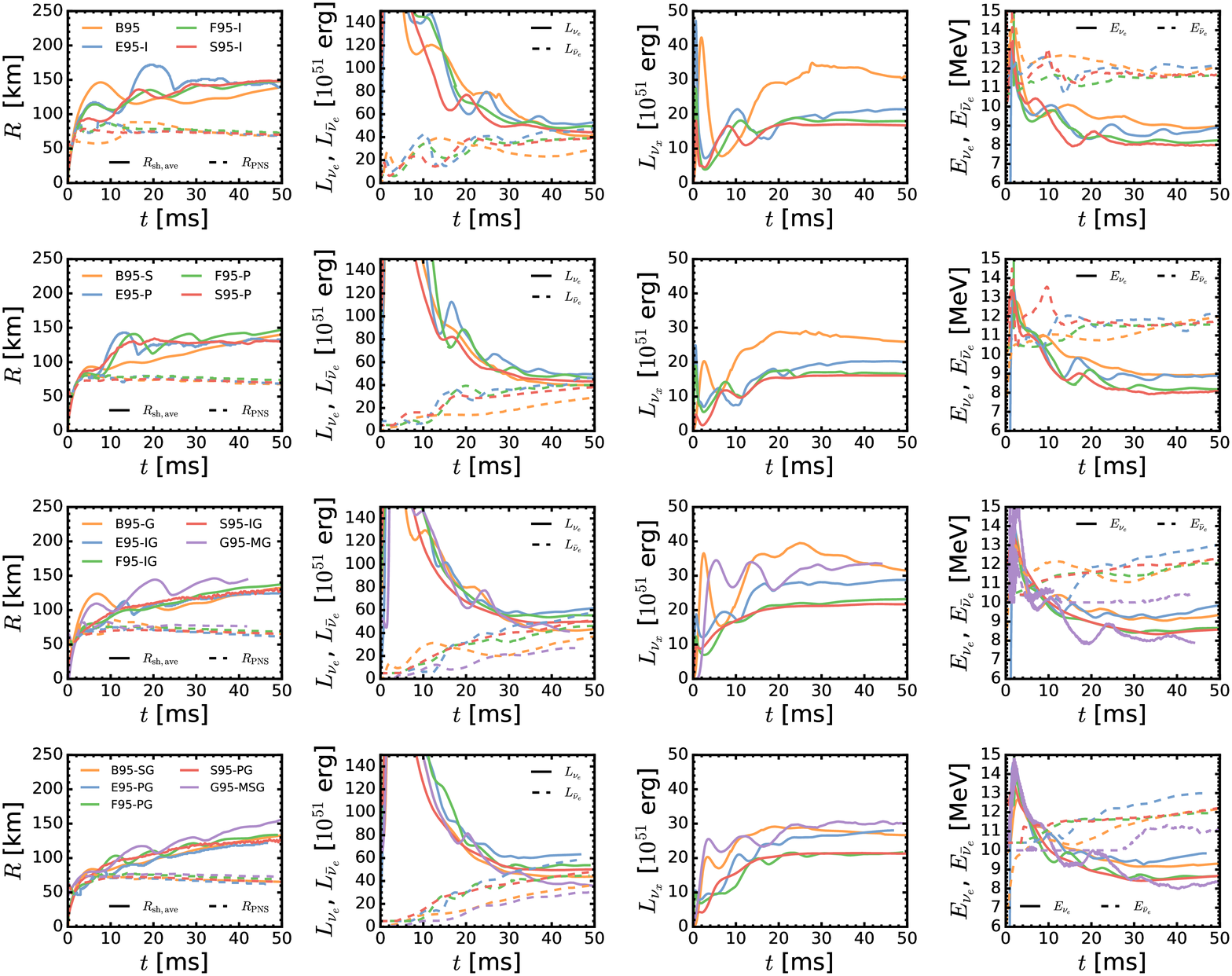}
  \caption{Time evolution of averaged shock radius, PNS radius, neutrino luminosities, and mean energies for different runs. Each row describes a comparison of different physics input as described in Table~\ref{table:overview}. Different color represents a simulation with a different hydrodynamics code.
\label{fig:overview}
}
\end{center}
\end{figure*}

A broad overview of the time evolution of several key quantities during
the post-bounce phase is shown in Fig.~\ref{fig:overview}. The first
row shows Newtonian runs with reduced input physics (labeled I in
table \ref{table:naming}). The second row shows Newtonian runs with
the effect of neutrino-electron scattering (labeled S or P in table
\ref{table:naming}). The third and fourth rows show the corresponding
general relativistic runs (labeled IG or MG with reduced physics and SG,
PG or MSG with neutrino-electron scattering). The panels in the left
column of the figure show the shock position as a function of time. All
four panels demonstrate that within $10$ ms the shock expands from a radius of a few kilometers to one of about $100$ km.

However, the four panels in the left column also show clear differences between the runs. We believe that
they are consistent with the following arguments: In the top panel it can be seen that run B95 reaches a
larger shock radius than the other codes, as well as in the third panel from above, B95-G and G95-MG reach a larger
shock radius than the other codes. This can be expected because in the IDSA, neutrinos are assumed to
propagate instantly through the stellar core due to the steady-state approximation for streaming neutrinos.
In the true transport codes B95-G and G95-MG this is not the case. There, the neutrinos require a few milliseconds to
move away from the location where they are produced. During these $5$-$10$ ms they block the phase space
for further neutrinos to be created. Hence, the shock in runs B95-G and G95-MG suffers less  from instantaneous neutrino
loss than the shock in the IDSA runs. This explanation is consistent with the fact that B95 also reaches  the
largest shock position in the topmost panel (there is no corresponding G95-M run). It is also consistent
with the observation that this difference in the early evolution of the shock radius is not seen in the
second and fourth panel from the top. These panels show graphs for runs that implement the effects of
neutrino-electron scattering. In these runs, the neutrino energy is lowered by down-scattering.
Thus, the neutrinos are emitted earlier. In this case, the codes produce a weaker shock from the beginning and a relatively
consistent shock expansion up to $10$ ms for all runs.

Further deviations appear at $\sim 15$ ms post-bounce. ELEPHANT produces
a broad hump in the top panel, which can also be seen in the FLASH data,
but to a much smaller extent. In the second panel from the top, both ELEPHANT
and FLASH show a similar hump at a slightly earlier time.
This is also
present in the third panel, albeit again smaller, but cannot be distinguished
in the fourth panel. We explain this hump with an early optimistic
expansion of matter supported by grid alignment effects. After this
first matter overturn, the alignment benefit vanishes and the shock
falls back to a more steady position.  In earlier, slowly rotating
models these humps are less visible (compare also to runs IR and PR in
Fig.~\ref{fig:overview_rot}).  In SPHYNX runs, the shock evolution shows
smaller humps or no humps at all. As SPH codes have no underlying
structured mesh, alignment effects are suppressed. Shock oscillations,
such as those present in runs S95-I and S95-P, can be linked to early prompt
convection.  After this first overshoot, the shocks in all models reach
consistent positions with deviations below $10\%$ between B95, E95,
F95 and S95.

The second column of the panels in Fig.~\ref{fig:overview} presents the luminosity as a function of time
for the electron flavor neutrinos (solid lines) and antineutrinos (dashed lines). After the peak luminosity
in the neutrino burst has occurred at $\sim 5$~ms after bounce, all luminosities decay to a value between
$4\times 10^{52}$~erg and $6\times 10^{52}$~erg at $50$~ms post-bounce.
At first glance, the neutrino luminosities seem to oscillate without
correlation during this decay.  Closer inspection reveals that there is a strong correlation between the dynamics of the shock front
discussed in the previous paragraph and the evolution of the electron
neutrino luminosity.  For example, run E95-I shows a prominent recession of
the shock at $24$~ms post-bounce. At exactly this time the corresponding
neutrino luminosity exhibits an equally prominent peak that is emitted by
electron capture when the hot matter behind the shock is compressed as it
slams into the surface of the PNS.  The same feature appears in runs E95-P and S95-P  at $16$~ms post-bounce, in run S95-I at 20~ms post-bounce, in run F95-P at $18$~ms post-bounce,
and in run G95-MG at $24$~ms post-bounce. Most features in the neutrino
luminosity can be explained in this way by the different dynamics of the
shock.

One exception is perhaps the luminosity of run B95: the prominent
recession of the shock at $12$~ms post-bounce would suggest a rather high
neutrino luminosity at that time.  The luminosity of B95 does show a
local maximum, but it does not exceed the luminosity of E95-I and F95-I. Here one
should additionally consider the PNS radius (dashed lines
in the leftmost panels), which initially shows a small PNS radius and a large shock radius. In this most optimistic run, a
substantial mass hovers between these radii and has not yet settled on
the PNS. At $26$~ms post-bounce, the situation is reversed:
as soon as the shock has receded after $14$~ms post-bounce
the PNS radius becomes even larger than in the other runs.
At this time, the luminosity of E95-I consistently exceeds the luminosities
of the other runs. The luminosity of run S95-I remains mostly below the
luminosities of the other runs because the shock expands slowly and exhibits
no recession except once around $20$~ms post-bounce.

The antineutrino luminosities depend much less on the accretion
rate. We find agreement of about $10\%$ between the runs that use the
IDSA and agreement after $\sim 20$~ms post-bounce between the two runs
that implement more sophisticated neutrino transport. However, the
IDSA-runs show consistently higher antineutrino luminosities than
the transport-runs.
We note that a similar deviation of the antineutrinos in the IDSA has been
noticed independently in [O'Connor et al. 2018], but not in [Liebendörfer
et al. 2009].  We also do not yet know why the antineutrino luminosities
calculated by the two transport codes B95 and fGR1 differ by about a
factor of two between $10$~ms and $20$~ms post-bounce. These questions
certainly merit further investigation and demonstrate how important such
comparisons are.

The mean energy of the neutrino flux, calculated by dividing the energy luminosity by the number luminosity,
is shown in the rightmost column of the panels.
In runs X95-I and X95-IG we find satisfactory agreement with deviations
of the order of $20\%$ during the most dynamic phase and of the order of $10\%$ after
the shock has assumed a more steady expansion. The longer-term neutrino
and antineutrino energies in ELEPHANT are consistently higher than in
FLASH and SPHYNX while the neutrino and antineutrino luminosities in fGR1
are significantly lower than in the other codes. The neutrino energies of
ELEPHANT agree well with those of AGILE-BOLTZTRAN in the runs without
neutrino-electron scattering and are higher than those of AGILE-BOLTZTRAN
in the runs that include neutrino-electron scattering. This is consistent
with the fact that with the IDSA the (anti-)neutrinos leave the star with
unchanged mean energy, while they are able to downscatter in the runs with
AGILE-BOLTZTRAN and fGR1, which both implement full neutrino-electron scattering.

We discover the most prominent differences between the codes in the luminosities
of the $\mu$- and $\tau$-neutrinos. The energy loss
due to the emission of heavy neutrinos in the Newtonian runs, which are
based on a gray leakage scheme for the heavy neutrinos, is smaller than the corresponding
energy loss calculated in AGILE-BOLTZTRAN with full neutrino transport by almost a
factor two.
Because the AGILE-BOLTZTRAN results assume similar values to those of an earlier
comparison \citep{liebendoerfer2005}, we assume that the AGILE-BOLTZTRAN data
are correct and  that the gray leakage scheme is too simple for the conditions that prevail
in Newtonian runs. This is consistent with the observation that the results
of fGR1 are closer to the results of AGILE-BOLTZTRAN than to the results
of the leakage scheme. In the more compact proto-neutron stars of the general
relativistic runs, where higher temperatures are reached, the leakage scheme
performs better and approaches the AGILE-BOLTZTRAN results to around
$20\%$ accuracy.

\begin{figure}
\centerline{\includegraphics[width=\linewidth]{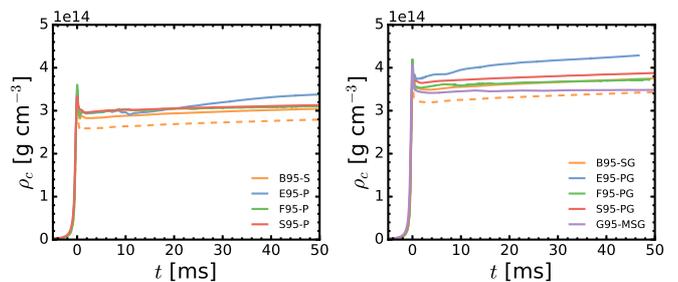}}
\caption{Central density evolution in all codes within a Newtonian (left) and GR (right) framework. The dashed line shows AGILE-BOLTZTRAN data before the extrapolation to the
center (cf. text).}
\label{fig:gr_rhoc}
\end{figure}

The differences in the energy loss rate by the emission of heavy neutrinos
points to the possibility that some differences in Fig.~\ref{fig:overview}
could be caused by a different evolution of the proto-neutron star.
Figure~\ref{fig:gr_rhoc} shows the central density as a function of time.
At first we note that the central density of AGILE-BOLTZTRAN (orange dashed line)
appears to have relatively low values. This is a result of the adaptive mesh,
which pulls grid points from the center of the PNS away to
the shock front. The center of the PNS is then less resolved
than in the other codes. As the innermost value in AGILE-BOLTZTRAN is a cell-centered
cell average, it is lower than the actual central density at the inner
edge of the cell. In order to cure this inconsistency in the comparison,
we extrapolate the
density profile of AGILE-BOLTZTRAN to the center of the computational domain using
the constraints that the density profile around the center be quadratic
with zero derivative at $r=0$. Furthermore, it must exactly reproduce the
density averages in the first and second cells given by the output files
of AGILE-BOLTZTRAN. After the extrapolation to the center, the density of AGILE-BOLTZTRAN
(orange solid line) shows better agreement with the central density of
the other codes.

\begin{figure}[htbp]
\begin{center}
 \includegraphics[width=\linewidth]{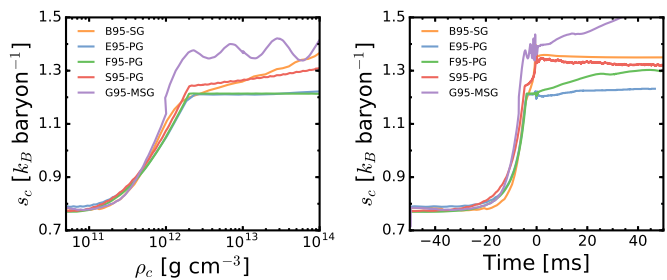}
   \caption{Central entropy as a function of central density (left) and as a function of time (right) for GR models including NES.
 \label{fig:entr_c}}
\end{center}
\end{figure}

In the Newtonian runs (left panel) we find an agreement to better than $5\%$
among all codes. However, the agreement in the general relativistic runs is
less perfect. E95-PG produces the most compact PNS with the
largest central density. G95-MSG produces the PNS with the
lowest central density. The central density depends on the self-gravity
of the neutron star, the central entropy and the central electron fraction.
We find good agreement with respect to the electron fraction. The $Y_e$ is
set by the parameterized deleptonization scheme and evolves in all models
in a similar fashion. Differences are found in the central entropy. Consistent
with the density differences, E95-PG shows the lowest entropy of $1.2$ $k_B$
per baryon while G95-MSG produces $1.4-1.5$ $k_B$ per baryon. Figure~\ref{fig:entr_c}
shows the evolution of the central entropy as a function of density during collapse (left panel) and as a function of time (right panel). The left panel shows 
that the main entropy difference is induced during the collapse phase when the density
crosses $\sim 10^{12}$ g/cm$^{-3}$. The right panel shows an unphysical slight
entropy increase after bounce in the multidimensional codes
that is most likely due to advection
of entropy from the shock-heated layers of the proto-neutron star towards the cold
center. This effect is not visible in the data of AGILE-BOLTZTRAN because the
core in spherical symmetry does not require substantial advection. It is also not visible
in the data of SPHYNX because advection is not an issue in the inherently
Lagrangian code. The effect is smaller in ELEPHANT than in the two
other 3D grid-based codes simply because the central entropy in ELEPHANT is
limited by the central entropy of AGILE-IDSA to mitigate this undesired central
entropy increase.

If we now go
back to Fig.~\ref{fig:overview} with these central entropy differences
in mind, we confirm in the lower left panel that
the PNS radius of G95-MSG is indeed larger than the one of
E95-PG. Also the shock radius of G95-MSG is larger than the one of E95-PG.
As expected, we find the inverse ordering in the neutrino luminosities: the
electron flavor luminosities of G95-MSG are significantly lower than those of
E95-PG. The difference of the mean energies between these two runs can be
explained by the same difference of the PNS structure: the
more compact PNS of E95-PG has neutrinospheres deeper in the
gravitational well and produces higher mean energies than the less compact
PNS of G95-MSG. 

In summary, we find a consistent overall evolution in all runs. Differences
during the collapse phase lead to slight differences in the entropy and
compactness of the PNS. The implementations of hydrodynamics
in our four codes optimize different aspects of the SN dynamics:
avoidance of grid-effects in SPH,
high resolution and efficiency at the shock front, adaptive capabilities and improved resolution
of the PNS, and last but not least, the possibility to include
full GR. These differences lead to deviations in the
details of the early shock expansion. Many differences in the evolution of
the neutrino luminosities can directly be linked to these hydrodynamic differences.
Some deviations, nota bene, require further investigation before we can extend
this comparison to later post-bounce times: Why are the electron
antineutrino luminosities produced by the IDSA rather large? And under which
PNS conditions is the leakage scheme for the $\mu-$ and $\tau-$neutrinos
sufficient - or insufficient?

\subsection{Neutrino electron scattering}
\label{nes}
Inelastic scattering of neutrinos off electrons plays an important role in helping neutrinos
to escape more freely from the center of the star as a consequence of down-scattering; it enhances deleptonization of the central core and thus leads to lower electron fraction $Y_e$.
The electron fraction is the most important quantity to determine the stability and the size of the inner core during stellar collapse.
The latter is crucial, since it defines the location of the shock at core bounce 
and affects the subsequent evolution of the neutrinospheres, which may be associated with the neutrino heating efficiency. 
The deleptonization process is mainly controlled by electron captures on free protons
and elastic/inelastic scattering of electron neutrinos.   
In this section, we compare its treatment used in each code and discuss their properties.

\begin{figure}[htbp]
\begin{center}
\includegraphics[width=\linewidth]{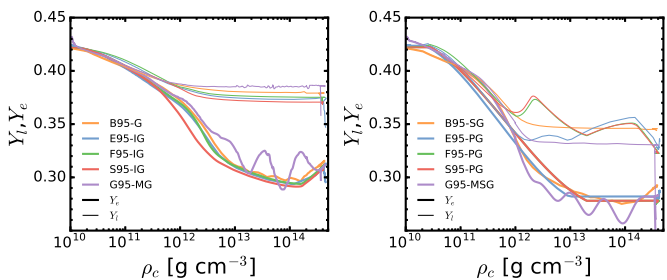}
  \caption{Central evolution of the electron (thick line) and lepton (thin line) fraction as a function of central
  density for various models.
  The left and right panels show this evolution without and with the NES effect being  taken into
account, respectively.
\label{KTf1}
}
\end{center}
\end{figure}

In Fig.~\ref{KTf1}, we plot the central evolution of the electron $Y_e$ (thick line) and total lepton $Y_l$ (thin) fractions as a function of central density for various models including GR.
  Left and right panels are without and with the NES effect, respectively.
  In the left panel, we see that the evolution of both $Y_e$ and $Y_l$ shows relatively good agreement among all models.
  Differences in both $Y_e$ and $Y_l$ are $\sim0.01$ at most.
  In addition, all numerical codes reproduce the neutrino trapping at nearly the same density of $\rho\sim10^{12}$ g cm$^{-3}$ and $Y_l$ stays nearly constant afterward.
  
\begin{table}
\begin{center}
\begin{tabular}{lccccc}
  & $\rho_1$ & $\rho_2$ & &  &  \\
Progenitor & (g~cm$^{-3}$) & (g~cm$^{-3}$) &$Y_1$&$Y_2$&$Y_c$\\
\hline
s15s7b2 & $3 \times 10^7$ & $2 \times 10^{13}$ & 0.5 & 0.278 & 0.035 \\
s15 & $2.2 \times 10^8$ & $9.5 \times 10^{12}$ & 0.5 & 0.279 & 0.022 \\
\end{tabular}
\end{center}
\caption{Summary of the parameter set for the parametrized deleptonization scheme. Model s15s7b2 \citep{woosley1995} is used for Newtonian and GR calculations. Model s15 \citep{woosley2007} for Newtonian calculations.}
\label{tab:PD}
\end{table}

\begin{figure}
\centerline{\includegraphics[width=\linewidth]{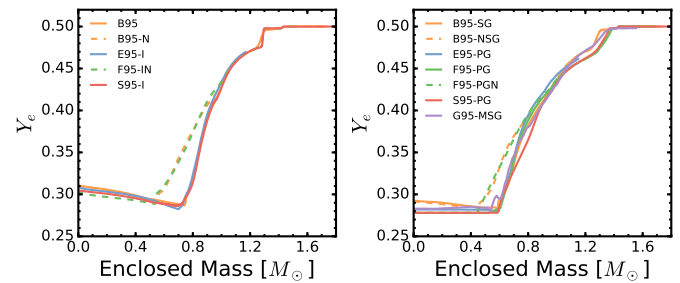}}
\caption{Influence of the neutrino stress in the $Y_e$ profiles at bounce. When neutrino stress is not included (dashed lines, i.e., N-labeled models) the inner core is $\sim 30\%$ smaller. The left panel is for IDSA Newtonian calculations, while the right panel is for GR calculations including neutrino-electron scattering.}
\label{fig:pd_nupres}
\end{figure}

In the panel on the right, we see the role of NES.
  Because of the downscattering of neutrinos off the electrons, lower-energy neutrinos escape more efficiently and the lepton fraction $Y_l$ becomes smaller compared to the model without NES.
  The parametrized deleptonization (PD) scheme enables $Y_e$ to be nearly an identical value with the one in B95-SG.
  In G95-MSG, the upper envelope of $Y_e$ shows a consistent value with the other models but with an oscillating behavior.
  Such oscillations are typical for simulations based on 12 energy groups. In corresponding spherically symmetric
  simulations \citep{Mezzacappa1993b,liebendoerfer2004} 
  these oscillations are explained by an insufficient resolution of the Fermi-energy of the degenerate electron neutrinos
  in the late collapse phase, which does not affect the overall evolution of the dynamics. In the
  present 3D study, however, we did not have the resources to prove this using a corresponding simulation
  with 20 energy groups.
  In models G95-MSG and B95-SG, neutrino trapping is appropriately reproduced which can be seen from the constant $Y_l$ evolution for $\rho\ga10^{12}$ g cm$^{-3}$. The value of
  $Y_l$ varies in models with PD scheme after neutrino trapping. This is because $Y_\nu$ is updated through the IDSA solver, but $Y_e$ is taken from the PD formulation. 
  Thus, the lepton number is not fully conserved with the PD scheme after neutrino trapping, but we consider that the overall picture is consistent with B95-SG/G95-MSG.

From the left panels of Fig.~\ref{KTf2}, the enclosed baryon mass with rest mass density larger than the neutrino trapping density ($\rho\sim10^{12}$ g cm$^{-3}$) is larger ($\sim0.95$ M$_\odot$) for models without NES (top-left panel)
than for those ($\sim0.9$ M$_\odot$) with NES (bottom-left panel).
This can also be seen in the second column of panels, where we can see that the enclosed baryon mass of the central core is $\sim0.05$ M$_{\odot}$ larger for models without NES ($\sim0.65$ M$_{\odot}$, top-right panel) compared to models with NES ($\sim0.60$ M$_{\odot}$, bottom-right panel).
This is because the additional pressure support from leptons leads to a more outward shock formation at bounce.
It should, however, be noted that the impact of this scattering process depends also on the electron capture rate. \cite{lentz2012b} reported that the down scattering can be blocked due to the production of lower-energy neutrinos when the up-to-date electron capture rate is used. Consequently, those models lead to the same results for the collapse phase, regardless of whether NES is included or not.

After bounce, we do not see any remarkable effects of NES.
Although the shock evolution and heavy lepton-type neutrino luminosity show minor differences during the first $\sim10$~ms after bounce (see bottom two rows in Fig.~\ref{fig:overview}) due to different shock position at bounce, the overall evolution shows no significant dependency on NES in time.
We, however, note that the lack of visible effects of NES in the post-bounce phase might be due to the shortness of our simulation times. \cite{just2018} recently reported, through their various models, that NES can turn a non-exploding model into a successful explosion. Nevertheless, such a noticeable effect appears at times ($t_{\rm pb}\ga 300$ ms) that are far beyond the end point of our simulations.

\subsubsection*{Calibration of parametrized deleptonization}
\label{sec:pd}

Before more modern rates for electron captures can be implemented in the IDSA,
the still missing implementation of NES in the IDSA would be important for the collapse phase.
In combination with traditional electron capture rates,
the inelastic scattering of neutrinos with electrons lowers the neutrino energy and accelerates deleptonization. 
In \cite{lentz2012, pan2016}, it has been shown that the central $Y_e$ at core bounce is
overestimated by about $10-15\%$ if NES is not taken into account during the collapse.
To effectively include NES, we follow the PD scheme described in \cite{liebendoerfer2005}, 
where the $Y_e$ is parametrized by the baryon density and the chemical potential. 
Table~\ref{tab:PD} summarizes the parameters that we used in this paper.
In Fig.~\ref{KTf2}, the PD scheme in X95-PG runs could lower the $Y_e$ from 0.31 in X95-IG to 0.28,
closely matching model G95-MSG, which includes NES.
  
In addition to electron fraction, neutrino stress is another important quantity that could affect the core size.
The contribution of neutrino pressure is not only included in the IDSA solver, 
but also implemented in the PD scheme \cite{liebendoerfer2005}.
The left panel of Fig.~\ref{fig:pd_nupres} shows a comparison of our 3D IDSA simulations
with AGILE-BOLTZTRAN simulations without NES (orange lines). 
The nearly identical electron fraction profiles at core bounce indicate that the IDSA 
could capture the deleptonization correctly when the NES is off.
Although the neutrino pressure contributes only a small fraction of the baryonic pressure, 
it could enlarge the inner core size by $\sim30\%$ (Fig.~\ref{fig:pd_nupres}). Moreover, the right panel of Fig.~\ref{fig:pd_nupres} demonstrates that the PD scheme could mimic the NES effects
and gives reasonable results compared to the AGILE-BOLTZTRAN simulation (the thick orange line; B95-SG).  
Furthermore, dashed lines present simulations of B95, F95-I, and F95-PG without the contribution from neutrino stress (with label `N'), 
leading to an inner core that is  smaller by $\sim30 \%$, consistent with the expected value.

\subsection{The role of general relativity}
\label{gr}
General relativity plays an essential role in CCSNe. The most
detailed 1D simulations to date, in which the GR Boltzmann
transport equation was solved in spherical symmetry, were performed by
\citet{wilson1971,mezzacappa1989,yamada1997,liebendoerfer2001,sumiyoshi2005,lentz2012},
and their 3D counterparts, with the multi-angle and multi-energy neutrino transport,
by \citet{sumiyoshi2012}.  Although it is not a full Boltzmann transport
scheme, the variable Eddington factor (VEF) method has been coupled to a
full GR hydro solver (e.g., \citealt{muller2012}, and references therein).
In this method, one can self-consistently determine the closure relation
from a model Boltzmann equation that is integrated to iteratively obtain
the solution up to the higher moments (i.e., the Eddington tensor)
until the system converges.

\citet{bruenn2001,liebendoerfer2001}, with baseline neutrino opacities, and \citet{lentz2012}, with the currently best available ones,
presented evidence that the average neutrino energy of any neutrino flavor during the shock reheating phase increases
when switching from Newtonian to GR hydrodynamics. They also pointed out that the increase is larger in magnitude compared
to the decrease due to redshift effects and gravitational time dilation. In addition, \citet{lentz2012} showed that
the omission of observer corrections in the transport equation lessens the triggering
of neutrino-driven explosions. In these fully fledged 1D simulations, a commonly observed disadvantageous aspect
of using GR to trigger neutrino-driven explosions is that the residency time of material
in the gain region becomes shorter due to the stronger gravitational pull. As a result of these competing ingredients, in the end
GR works against the triggering of the neutrino-driven explosions in 1D. In fact, the maximum shock extent in the post-bounce
phase is shown to be $\sim$20\% smaller when switching from Newtonian to GR hydrodynamics \citep{bruenn2001,liebendoerfer2001,sumiyoshi2005,lentz2012}.

From Fig.~\ref{fig:overview}, we can see that the mean PNS radii are $\sim10$ \% smaller in (effective-)GR models
compared to Newtonian ones (see, e.g., the second and fourth panels from the top in the leftmost column), confirming the same effect also in 3D simulations.
GR also affects the shock evolution. Irrespective of NES, the shock front appears $\sim10$ km more inward in (effective-)GR
models.
Furthermore, because of the more compact and hotter PNS, neutrino luminosities in all flavors become $\sim20$ \% higher. AGILE-BOLTZTRAN takes into account both the gravitational redshift and Doppler shift terms.
Therefore, we can explore the influence of GR on the emergent neutrino energies by comparing B95-S and B95-SG.
As a consequence of the competition between the gravitational red-, Doppler-shift, and hotter PNS surface,
the mean neutrino energies are higher   in B95-SG than in B95-S by $\sim0.5$ MeV.
This simply reflects the effect of GR, leading to a more compact PNS core,
which can also be confirmed in Fig.~\ref{fig:gr_rhoc}. This figure shows a comparison of the central density evolution between Newtonian (left) and GR (right) models.
In GR models, the central density shows values that are systematically higher by $\sim 20-30\%$. To see if these GR effects work advantageously on the shock revival, we need to perform simulations that cover
more of the post-bounce phase.

\begin{figure*}[h!]
\begin{center}
\includegraphics[width=\linewidth]{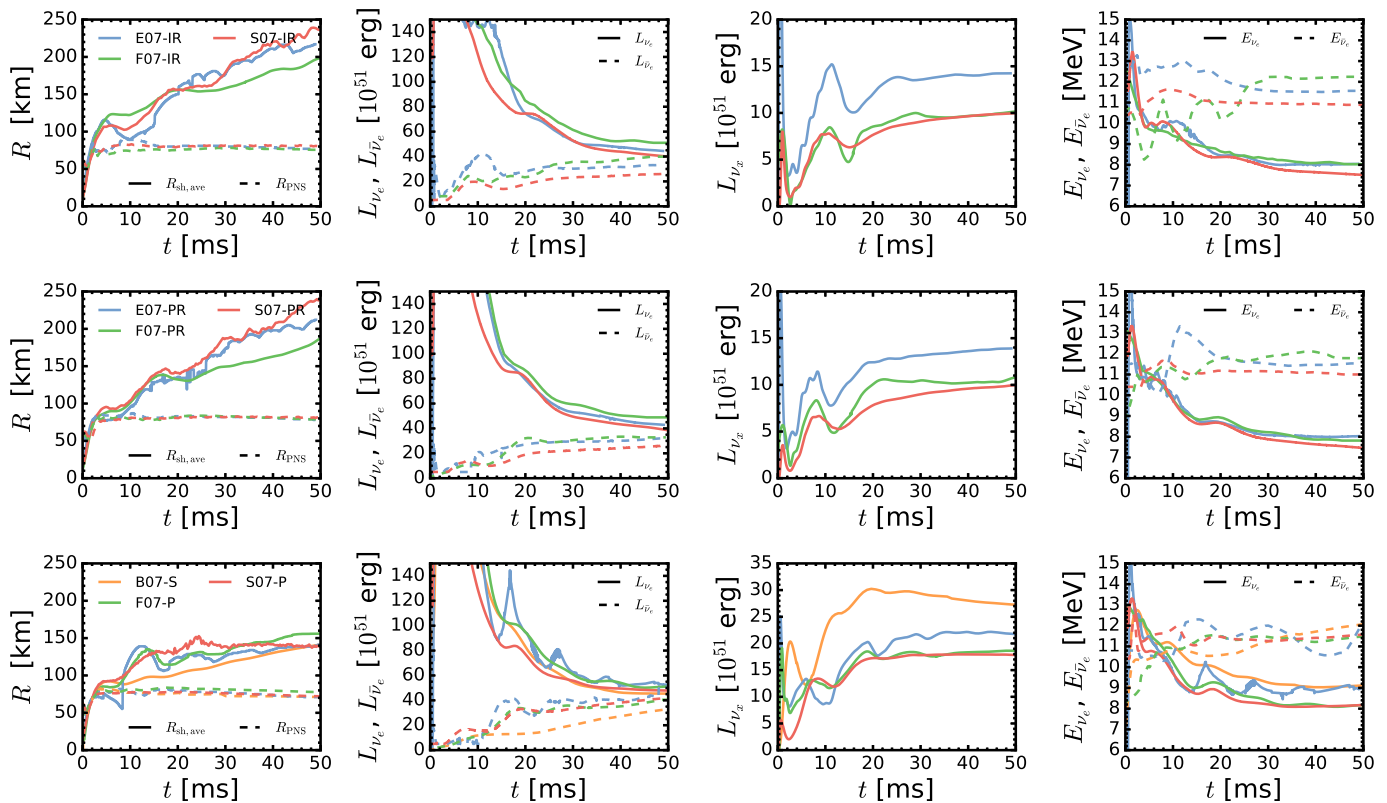}
  \caption{Similar to Fig.~\ref{fig:overview} but for rotating models X07-IR (first row) and X07-PR (second row). Non-rotating models X07-P (third row) are shown here for comparison purposes.
  \label{fig:overview_rot}
}
\end{center}
\end{figure*}

\begin{figure}[b]
\centerline{\includegraphics[width=\linewidth]{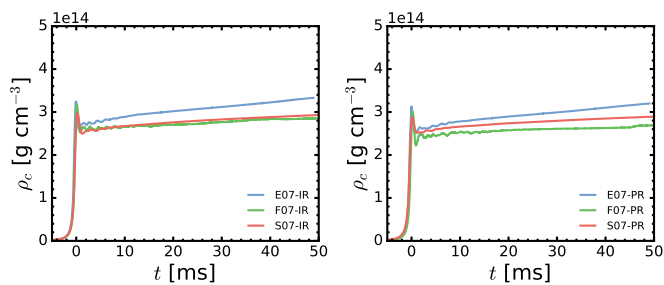}}
\caption{Central density evolution for rotating models without (left) and with (right) parametrized deleptonization.}
\label{fig:rhoc_rot}
\end{figure}

\begin{figure*}
\centerline{\includegraphics[width=\linewidth]{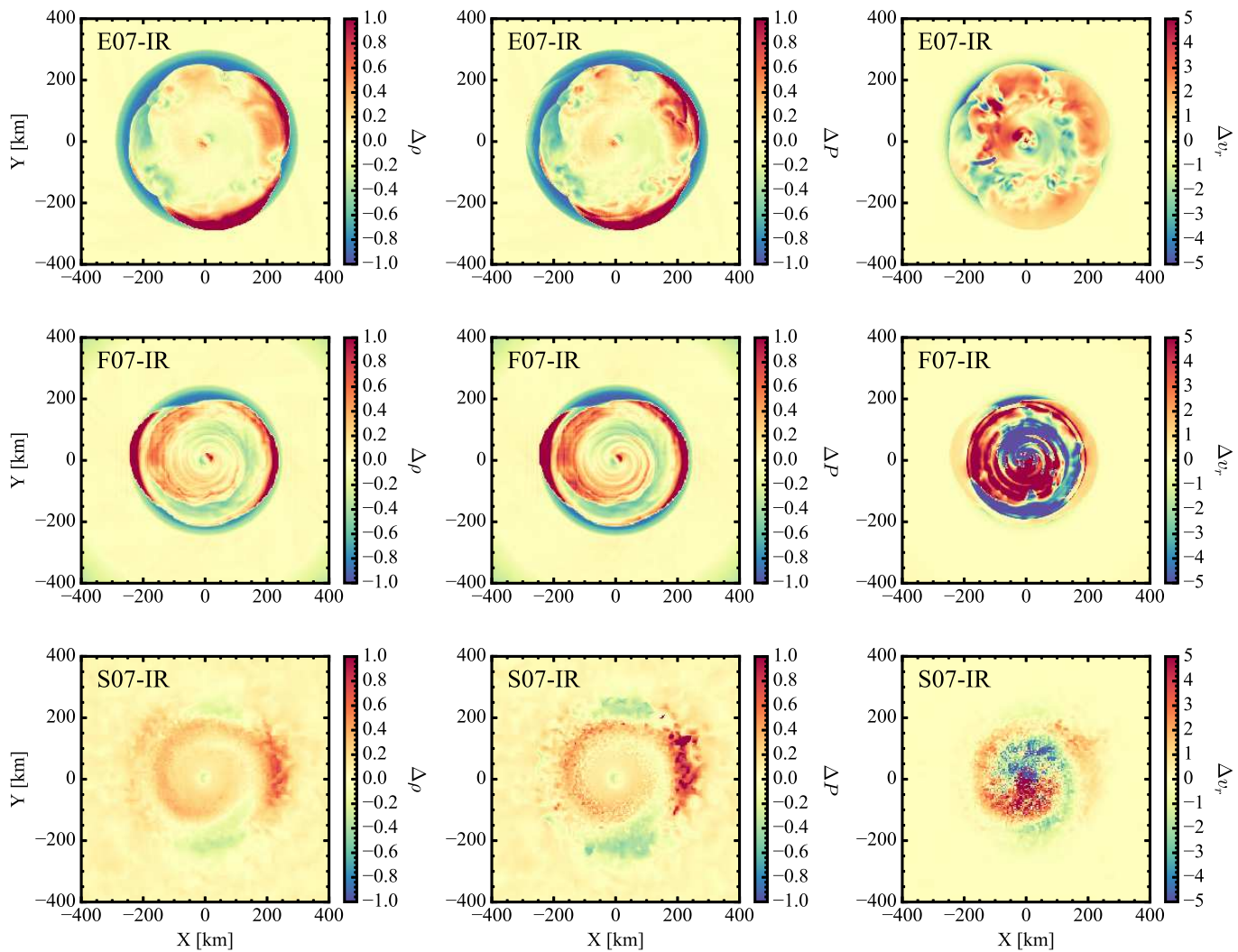}}
\caption{Deviations from the spherical average of density (left), pressure (center), and radial velocity (right) profiles for rotational models X07-IR at 50 ms.}
\label{fig:rotation}
\end{figure*}

\begin{figure}
\centerline{\includegraphics[width=\linewidth]{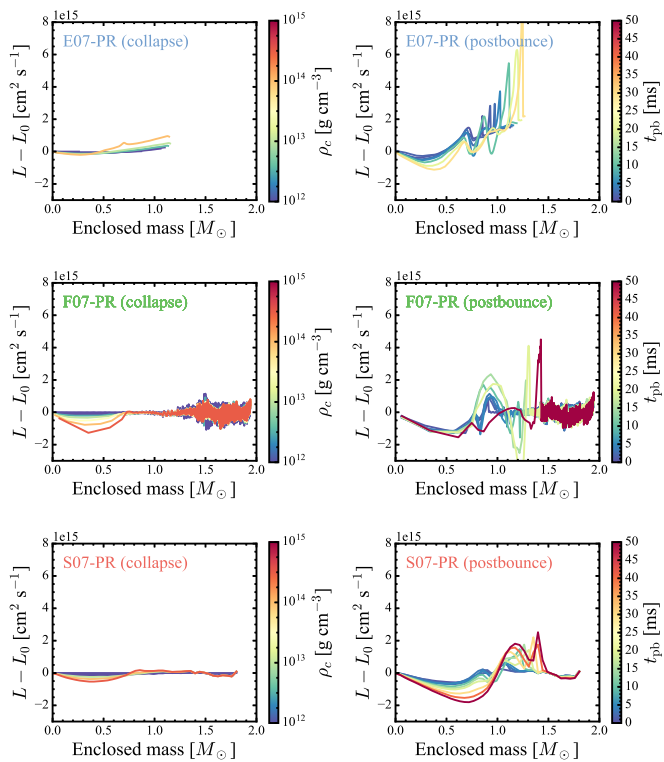}}
\caption{Specific angular momentum differences with respect to the initial specific angular momentum distribution at different central densities during collapse (left panels) and at different post-bounce times (right panels). From top to bottom, the represented simulations correspond to ELEPHANT, FLASH, and SPHYNX codes.} 
\label{fig:spangmom_diff}
\end{figure}

\subsection{The role of rotation}
\label{rotation}
With the advent of multidimensional SN simulations
the role of rotation is gaining increased attention from the community
\citep{marek2009,suwa2010,yokozawa2015,takiwaki2016,janka2016,gilkis2018,summa2018},
as it can have sizable effects on the emergence of
an explosion as well as on important SN observables. For example,
gravitational waves produced from the development of triaxial
instabilities triggered by rotation will carry information about the
supernuclear EOS that is encoded in their waveform. From these, we could
impose constraints on rotation rates of the iron core and the PNS, which are not well known.

Following the setup presented in \cite{yokozawa2015} we implemented a shellular rotation \citep{eriguchi1985} assigning an angular velocity $\Omega$ at each cylindrical radius $r$ in the form of

\begin{equation}
\Omega(r)=\frac{\Omega_0 r_0^2}{r^2+r_0^2},
\label{shellular}
\end{equation}

\noindent
where $\Omega_0$ is the angular velocity at the center and $r_0$ is a characteristic distance that controls the amount of differential rotation. For the limit of small values of $r_0$, Eq.~(\ref{shellular}) approaches a configuration with constant specific angular momentum, while for large values it tends to a rigidly rotating model. We chose $r_0=1000$~km and $\Omega_0=\pi$~rad~\inv{s}. The former ensures a nearly rigid rotation of the core of the star, while the latter corresponds to a spin period of 2~s. We note that this is  roughly
a factor 50 more rapid than the predictions by detailed stellar evolution models \citep{heger2005} that include angular momentum transfer by magnetic processes. Nevertheless, many publications employ such a rapid rotation profile, even if it is probably only compatible with a differentially rotating core in hypernovae \citep{burrows2007} and/or a fast rotating black hole at the origin of gamma-ray bursts \citep{macfadyen1999}. In order to be able to compare with the literature we kept the same rotational profile. We also highlight here that Eq.~\ref{shellular} generates a rotational profile in cylindrical shells, which was not applicable in ELEPHANT. As explained in Sect.~\ref{elephant}, this code simulates the inner region of the star with a 3D Cartesian mesh while its boundary conditions are provided by a simultaneous calculation with the 1D code AGILE-IDSA. This leads us to impose only spherical rotational profiles for the 3D inner region. Therefore, ELEPHANT applies Eq.~\ref{shellular} with $r$ and $r_0$ representing spherical radii instead of cylindrical radii. As a consequence, the initial rotational profile of ELEPHANT is different from those of SPHYNX and FLASH as it rotates more slowly ($\sim 6\%$).

Figure~\ref{fig:overview_rot} presents an overview of the average shock position and PNS radius (left column), electron-neutrino luminosity (second column), heavy-lepton neutrino luminosity (third column), and mean energies for all neutrino species (last column) for the calculated rotating models without (IR) and with (PR) parametrized deleptonization. In order to clearly see the influence of rotation, we added a third row to Fig.~\ref{fig:overview_rot} with the results of X07-P models, which include the same physics as the rotating models but without rotation. Overall, there is good agreement among the codes for the rotating models. All three do not show signs of a rapidly stalled shock within the simulated time, in contrast to the non-rotating models where the shock halts at about 100-150 km within the first 20~ms. As expected, rotation helps the expansion of the shock due to the deposition of angular momentum in outer layers \citep{fryer2004,suwa2010,nakamura2014}. There is good agreement in the neutrino and anti-neutrino luminosities among all thee codes, although SPHYNX tends to be slightly less luminous than the other two codes for anti-neutrinos (second column), which correlates with the lower mean energy deposited in that neutrino species (fourth column). Nevertheless, the main difference among codes can be seen in the heavy-lepton neutrino luminosity (third column), where SPHYNX and FLASH show good agreement, but ELEPHANT produces $\sim 40$\% more luminosity. The reason for this is the different rotational profile of ELEPHANT, which produces a more compact PNS, consistent with its central density evolution shown in Fig.~\ref{fig:rhoc_rot}. As expected, ELEPHANT simulations tend to form more compact objects than those of FLASH. SPHYNX obtains intermediate results, where the biggest differences are $\sim 15\%$.

The angular momentum transfer is crucial in rotating scenarios. We note that the angular momentum transport is strongly enhanced after bounce, where the shock launch aids the transfer of fast-rotating matter to larger radii, driving a rapid decrease of the angular momentum of the inner core, as it is transported to the surface of the neutron star. This mechanism excites the $m=1$ spiral-arm modes, leaving a signal in the fluid that we can see in Figs.~\ref{fig:rotation} (for IR models). In this latter figure, at $t_{pb}=50$~ms, we plot the deviation from the spherical average of the density (left), pressure (center), and radial velocity (right) in a thin slice of the 3D domain along the equatorial plane. We can see in all three codes that a spiral, high-density, high-pressure wave is moving outwards from the center of the system. SPHYNX and FLASH produce very similar results, both in magnitude and size of the spiral perturbation at the same time. Due to the different rotational profile, ELEPHANT results are not directly comparable to those of the other two codes. Nevertheless, a spiral shape is still visible on all three magnitudes, albeit $\sim 15 \%$ more extended. Our results are similar to those of \cite{takiwaki2016} (see their Fig.~3).

Additionally, we can also follow the transfer of angular momentum directly during the whole evolution, plotting the difference of specific angular momentum with respect the initial specific angular momentum distribution as a function of the enclosed mass. This is represented in Fig.~\ref{fig:spangmom_diff} for ELEPHANT, FLASH, and SPHYNX (from top to bottom) at different stages of the collapse (left column) and at different times after bounce (right column). If we focus on the first column, we can see that even during collapse, some angular momentum is lost in the inner core for all three codes. This may be due to two factors that occur at the same time: the appearance of shear and the lack of angular momentum conservation by the code. Some shear might be expected during the collapse of a rotating star due to the stronger centrifugal forces experienced by the material with higher angular momentum (i.e., in the equatorial plane). As a consequence, its collapse proceeds slower than material with lower angular momentum (e.g., in the polar regions) and a certain quantity of shear appears, causing angular momentum to pile up at larger radii (this is especially visible at $t_{pb}=0$~ms in the SPHYNX run, in the small bump from M~$\sim 0.8$\msun up to $\sim 1.3$\msun in Fig.~\ref{fig:spangmom_diff}). Additionally, some angular momentum is lost due to numerical diffusion. This is more relevant in the mesh-based codes ELEPHANT and FLASH, while SPHYNX does conserve it better. This is an expected behavior of an SPH code, which, being purely Lagrangian, is specially constructed to conserve momentum and energy. It is worth noting though, that some angular momentum is also lost in SPHYNX due to the calculation of the gravitational force, which, to some extent, always breaks the perfect conservation properties of the SPH formalism. We can estimate the total angular momentum that is lost due to numerics and shear by integrating the losses and gains of the specific angular momentum as separate functions of the enclosed mass. If angular momentum is perfectly conserved, both magnitudes should be equal and therefore their difference should be zero. We found that, up to bounce, SPHYNX has a losses-to-gain difference below 1\%, while the value for FLASH is below 3\%, and the value for ELEPHANT is below 4\%. This percentage gives an upper limit to the angular momentum losses at the (low) spatial resolution at which these simulations have been performed. Increasing the resolution, this ratio improves considerably for all three codes.
Unfortunately we cannot clearly disentangle the undesired angular momentum loss on the scale of the dimensions of the proto-neutron star from unavoidable angular momentum loss due to dissipation of turbulence on smaller scales.
But in our investigation of the collapse phase and bounce we consider the loss-to-gain difference as a reasonable estimate for undesired angular momentum losses because our simulations do not develop turbulence during the very short
collapse of our perfectly spherically symmetric progenitor models.

After bounce (second column) it is clear that angular momentum is extracted from the inner core and transported to the outer layers. 
The time and the regions at which angular momentum changes differ among codes.
The presence of convection is the reason for the complex structure of these profiles. The different treatments of dissipative terms in the different codes and the low spatial resolution may also have an important impact. Despite these
quantitative differences, it can be seen in all three codes how angular momentum moves up along the enclosed mass axis, and, in fact, reaches similar enclosed masses, that is, $\sim 1.4$~\msun, in FLASH and SPHYNX at about the same time ($t\sim 50$~ms).

\subsection{Computational resources}
\label{resources}

Using different codes with different levels of approximation implies great differences in computational cost, which, in turn, affect the spatial resolution that we can afford.  In Table~\ref{table:cpuh} we provide some details regarding the CPUh consumption, parallelization methods, number of fluid elements, and resolution of each code in this work in order to give some idea of the computational costs. Some of the calculations were done in different facilities, so the used resources presented in Table~\ref{table:cpuh} are merely indicative. Additionally, the computational efficiency of each code is obviously highly dependent on its parallelization methods and its implementation. Hence, the values presented in Table~\ref{table:cpuh} reflect the current state of development of these codes.

ELEPHANT provides (highly efficient) simulations with constant spatial resolution (1~km), which is very good at the shock position but coarse at the PNS. This is mitigated in FLASH via the usage of adaptive mesh refinement (AMR). The resolution at the PNS is greatly improved and a third-order scheme is used at the price of increased computational resources. fGR1 uses nested meshes of finer resolution, so there is no AMR overhead. Nevertheless, the inclusion of full GR and M1 treatments requires considerable computational resources. On the other hand, SPHYNX reaches similar resolution in the PNS with considerably less fluid elements, but, as in SPH codes resolution follows density, more diluted regions like that of the shock are much coarser. In the case of ELEPHANT and SPHYNX, the lack of resolution in some regions of the core collapse scenario can be compensated with higher amounts of fluid elements. Subsequently, the viability of the simulation relies on its computational efficiency and the scaling of the code in a parallel implementation. It is clear that the use of one code over another can be beneficial depending on the focus of the study. For example, increasing the amount of SPH particles up to more `production-simulation-like' amounts (e.g., $2\times 10^7$) will increase the spatial resolution by a factor $\sim 4.6$ reaching $\sim 100$~m resolution in the PNS. On the other hand, ELEPHANT can provide a very high resolution ($\sim 500$~m) in the gain region and at the same time include magnetic fields.

\begin{table*}[h!]
\begin{center}
\begin{tabular}{|c|c|c|c|c|c|c|}
\hline
Code & Type & Discretization & \# fluid elements & Parallelization & CPUh & Max. resolution \\
 &  &  &  &  &  & (m) \\
\hline\hline
ELEPHANT & Eulerian & Cartesian mesh & $2.1\times10^8$ & MPI + GPU & 79,000 & 1,000 \\ 
FLASH & Eulerian & AMR & $1.9\times10^7$ & MPI + GPU & 117,000 & 488 \\ 
SPHYNX & Lagrangian & SPH particles & $2\times10^5$ & MPI + OpenMP & 13,000 & 477 \\
fGR1 & Eulerian & Nested mesh & $3.5\times10^6$ & MPI & 245,000 & 458 \\ 
\hline
\end{tabular}
\end{center}
\caption{Summary of the details and parallelization methods in each code, an estimation of the CPUh to calculate collapse and post-bounce until 50 ms for models X95-PG (ELEPHANT, FLASH, and SPHYNX) and model G95-MSG (fGR1), and maximal spatial resolution at $t=50$~ms.}
\label{table:cpuh}
\end{table*} 

\section{Conclusions}
\label{conclusions}

In this work, we studied the same core collapse scenario in 3D with four different hydrodynamics codes including, for the first time, Eulerian and Lagrangian codes in the same study. In addition, we also compared the four hydrodynamics codes with a spherically symmetric code with Boltzmann transport. Within this framework, we varied three parameters in order to explore their different impacts on different hydrodynamics codes, namely the neutrino-electron scattering, a monopole GR correction, and rotation. Additionally, we discussed the differences between the approximate treatments and more detailed calculations, that is, IDSA + Parametrized deleptonization versus M1 neutrino transport, and the monopolar GR correction versus~full GR.

The comparison of different neutrino transport methods in a SN simulation
revealed that all our codes are able to reproduce a reasonable overall evolution of the early post-bounce
phase in 3D. The strong coupling and
feedback between the proto-neutron star compactness, shock dynamics, neutrino luminosities, and mean energies is
in all cases consistent. Despite their different discretization methods of the Euler equations, including radically different domain decompositions, parallelization approaches, and the use of static or nested meshes, AMR or SPH, the deviations of the results are of the order of
$10\%$, and sometimes $20\%$. This comparison leads us to propose a requirement for further investigation of the luminosities
of antineutrinos in the IDSA, which tend to be rather high. Furthermore, the leakage scheme for the heavy neutrinos
may not accurately represent the corresponding energy loss under all conditions of the proto-neutron star.
Here, a better scheme would be desirable (e.g., \citealt{perego2014,takiwaki2014}). The most computationally expensive full GR code still
seems to struggle with resolution and performance issues. The whole ensemble of codes however presents many virtues
if referred to in combination: excellent resolution of the proto-neutron star and angular momentum conservation (SPHYNX),
flexibility and open accessibility for efficient 3D SN simulations with GPU-acceleration (FLASH),
superior shock resolution and efficient inclusion of magnetic fields (ELEPHANT), fully GR
hydrodynamics and neutrino transport (fGR1). The results presented here verify the usefulness of these four codes with respect to their coupling with the neutrino treatment and to the implementation of gravity as previously discussed.

We also checked the influence of neutrino-electron scattering, which plays a significant role during collapse phase.
Whether we implement the inelastic scattering kernel explicitly in the source terms 
or control the deleptonization process in a parametrized way, we confirm that the lepton profiles within the inner core show converged results among all models. 
Furthermore, the hydrodynamics and neutrino profiles post-bounce show consistent evolution with deviations of
$\sim$10\%,
which is an acceptable level considering that we use totally independent codes.

The effects of GR are investigated by comparing the Newtonian and (effective-)GR models.
Due to the stronger gravitational force, we see more compact PNS cores and shock positions in GR models.
This is consistent with previous studies \citep{bruenn2001,liebendoerfer2001,sumiyoshi2005,lentz2012}.
From hotter and more compact PNS cores, the emergent neutrinos show systematically higher
neutrino energies and luminosities in all flavors,
even after taking into account the effect of gravitational redshift (e.g., comparing B95-S and B95-SG).
Within our simulation times, we do not see any remarkable difference between full- and effective-GR models, and therefore
we consider the usage of an effective-GR potential as justified.
However, our simulation time is short and longer simulations
are required to see for how far the effective-GR potential works properly and to determine the time when those two methods diverge.
Furthermore, it would also be interesting to check under which conditions GR acts advantageously on the shock revival compared to the Newtonian case.

The inclusion of rotation helps the expansion of the shock in general. Even when using different rotation profiles (ELEPHANT vs. FLASH and SPHYNX), the shock position at 50~ms can be $26-70$\% further than in models without rotation. We confirmed the presence of a $m=1$ spiral perturbation that drives matter from the inner core into the gain region, aiding the explosion. This is partially compensated by a less dense and cooler PNS, but the net balance between both effects favors the expansion of the sock. Additionally, we could evaluate the performance of the different hydrodynamics codes with respect to the angular momentum transport. With the present (low) resolution in this study the angular momentum conservation was better than 4\% .

Additionally, we were able to clearly demonstrate that an SPH code with a spectral neutrino treatment produces results comparable to those of Eulerian codes. Interestingly, this opens the possibility to consider SPH as a competitive, efficient method to simulate core-collapse SNe with adaptive resolution and open boundaries. This is particularly useful when one aims to study the PNS at very high resolution, when angular momentum conservation is capital, or when the PNS kick is the focus of an investigation. We note, that there is remarkable agreement between the results of SPHYNX and FLASH in many simulated models. In particular when the effective GR potential is used. Both codes provide a high spatial resolution at the PNS, which is important for an accurate compactness of the proto-neutron star and thus the overall evolution of the system.

It should be noticed that there are several multi-dimensional effects that deserve further investigation, such as turbulence, standing accretion shock instability, magnetic fields, and gravitational wave emission.

Additionally, extending this kind of comparison to other progenitors and longer timescales is of utmost importance.
However, this kind of simulation is extremely costly in terms of computational resources and as a consequence has been relegated. We think, though, that the time to perform detailed comparisons among complex production codes and across many different simulations is getting closer. It is in this spirit that our work provides the first outlook on SN code comparisons in 3D.

\section*{Acknowledgements}
The authors want to thank S. C. Whitehouse for his extensive work on ELEPHANT.
This work has been supported by the European Research Council (FP7) under ERC Advanced Grant Agreement No. 321263 - FISH,
by the Swiss Platform for Advanced Scientific Computing (PASC) project DIAPHANE (RC and KCP) and SPH-EXA (RC),
by ERC Starting Grant EUROPIUM-677912 (TK),
and by JSPS KAKENHI Grant Number JP17H01130 and JP17H06364 (TK).
A part of numerical computations were carried out on Cray XC30 at CfCA, National Astronomical Observatory of Japan (TK).
The authors acknowledge the support of sciCORE (http://scicore.unibas.ch/) scientific computing core facility at University of Basel, 
the Swiss National Supercomputing Center CSCS (http://cscs.ch) under grant No.~661 and No.~840, 
where these calculations were performed. 
{\tt FLASH} was in part developed by the DOE NNSA-ASC OASCR Flash Center at the University of Chicago.
Analysis and visualization of the FLASH simulation data were completed using the {\tt yt} analysis toolkit \citep{turk2011}.

\bibliographystyle{aa}
\bibliography{bibliography}

\end{document}